\documentclass[format=acmsmall, review=false, screen=true]{acmart}

\usepackage{booktabs} 
\usepackage{url}
\usepackage{amsfonts,amsmath,amssymb}
\usepackage[vlined,linesnumbered,tworuled,algo2e]{algorithm2e}
\usepackage[caption=false]{subfig}
\usepackage{tikz}
\usepackage{ifthen}
\usepackage{wrapfig}
\usepackage{algpseudocode}
\usepackage{todonotes}
\usepackage{color}
\usepackage{xspace}
\usetikzlibrary{arrows} 

\SetAlFnt{\small}
\SetAlCapFnt{\small}
\SetAlCapNameFnt{\small}
\SetAlCapHSkip{0pt}
\IncMargin{-\parindent}

\acmJournal{JEA}
\acmVolume{X}
\acmNumber{X}
\acmArticle{X}
\acmYear{2017}
\acmMonth{11}
\copyrightyear{2017}

\received{X}

\newcommand{\MRI}{{\textsc{MRI}}\xspace}
\newcommand{\MBI}{{\textsc{MBI}}\xspace}

\newcommand{\MSC}{{\textsc{MSC}}\xspace}
\newcommand{\XC}{{\textsc{X3C}}\xspace}
\newcommand{\Greedy}{{\textsc{Greedy}}\xspace}
\newcommand{\TopKbet}{{\textsc{Top-k}}\xspace}
\newcommand{\etal}{et al.\xspace}

\newcommand{\optproblem}[4]{
\begin{center}
\begin{tabular}{lp{0.83\textwidth}}
\hline
\multicolumn{2}{c}{\textbf{#1}}\\
\hline
\textbf{Given:} &  #2\\
\textbf{Solution:} & #3\\
\textbf{Objective:} & #4\\
\hline
\end{tabular}
\end{center}
}

\makeatletter
\newcommand\footnoteref[1]{\protected@xdef\@thefnmark{\ref{#1}}\@footnotemark}
\makeatother

\begin{document}
\title{Improving the Betweenness Centrality of a Node by Adding Links}

\author{Elisabetta Bergamini}
\affiliation{%
  \institution{Karlsruhe Institute of Technology}
  \streetaddress{Am Fasanengarten 5}
  \city{Karlsruhe}
  \postcode{76161}
  \country{Germany}}
\email{elisabetta.bergamini@kit.edu}

\author{Pierluigi Crescenzi}
\affiliation{%
  \institution{University of Florence}
   \streetaddress{Viale Morgagni 65}
   \postcode{50134}
  \city{Florence}
  \country{Italy}}
  \email{pierluigi.crescenzi@unifi.it}

\author{Gianlorenzo D'Angelo}
\affiliation{%
  \institution{Gran Sasso Science Institute (GSSI)}
   \streetaddress{Viale F. Crispi, 7}
   \postcode{67100}
  \city{L'Aquila}
  \country{Italy}}
  \email{gianlorenzo.dangelo@gssi.infn.it}
  
\author{Henning Meyerhenke}
\affiliation{%
  \institution{Institute of Computer Science, University of Cologne}
  \streetaddress{Weyertal 121}
  \city{Cologne}
  \postcode{50931}
  \country{Germany}}
\email{h.meyerhenke@uni-koeln.de}
  
 \author{Lorenzo Severini}
\affiliation{%
  \institution{ISI Foundation}
   \streetaddress{Via Chisola, 5}
   \postcode{10126}
  \city{Torino}
  \country{Italy}}
  \email{lorenzo.severini@isi.it}
  
  \author{Yllka Velaj}
\affiliation{%
  \institution{University of Chieti-Pescara}
   \streetaddress{Viale Pindaro 42}
   \postcode{65127}
  \city{Pescara}
  \country{Italy}}
  \email{yllka.velaj@gssi.infn.it}
  
            
\begin{abstract} 
Betweenness is a well-known centrality measure that ranks the nodes according to their participation in the shortest paths of a network. In several scenarios, having a high betweenness can have a positive impact on the node itself. Hence, in this paper we consider the problem of determining how much a vertex can increase its centrality by creating a limited amount of new edges incident to it. In particular, we study the problem of maximizing the betweenness score of a given node -- Maximum Betweenness Improvement (MBI) -- and that of maximizing the ranking of a given node -- Maximum Ranking Improvement (MRI). We show that MBI cannot be approximated in polynomial-time within a factor $(1-\frac{1}{2e})$ and that MRI does not admit any polynomial-time constant factor approximation algorithm, both unless $P=NP$. We then propose a simple greedy approximation algorithm for MBI with an almost tight approximation ratio and we test its performance on several real-world networks. We experimentally show that our algorithm highly increases both the betweenness score and the ranking of a given node and that it outperforms several competitive baselines. To speed up the computation of our greedy algorithm, we also propose a new dynamic algorithm for updating the betweenness of one node after an edge insertion, which might be of independent interest. Using the dynamic algorithm, we are now able to compute an approximation of MBI on networks with up to $10^5$ edges in most cases in a matter of seconds or a few minutes.
\end{abstract}


\maketitle
\newpage
\section{Introduction}
In recent years, the analysis of complex networks has become an extremely active research area. One of the main tasks in network analysis is computing the ranking of nodes based on their structural importance. Since the notion of importance can vary significantly depending on the application, several \textit{centrality measures} have been introduced in the literature. One of the most popular measures is \textit{betweenness centrality}, which ranks the nodes according to their participation in the shortest paths between other node pairs. Intuitively, betweenness measures a node's influence on the flow circulating through the network, under the assumption that the flow follows shortest paths. 

Computing betweenness centrality in unweighted graphs requires $\Theta(n m)$ time with Brandes's algorithm~\cite{Brandes01betweennessCentrality}, where $n$ is the number of nodes and $m$ is the number of edges. Since this can be prohibitive for very large networks, several approximation algorithms exist in the literature~\cite{DBLP:journals/corr/BorassiN16,DBLP:conf/alenex/GeisbergerSS08,DBLP:journals/corr/RiondatoU16,DBLP:journals/datamine/RiondatoK16}. Also for dynamic networks that evolve over time, such as social networks and the Web graph, recomputing betweenness at every time step can be too expensive. For this reason, a variety of dynamic algorithms have been proposed over the last years~\cite{DBLP:conf/socialcom/GreenMB12,DBLP:conf/asunam/KasWCC13,DBLP:journals/isci/LeeCC16,DBLP:conf/mfcs/NasrePR14,DBLP:conf/isaac/PontecorviR15,DBLP:conf/sea/BergaminiMOS17}. These algorithms usually keep track of the betweenness scores and additional information, such as the pairwise distances, and update them accordingly after a modification in the graph.
Another problem that has recently been considered for betweenness and other centrality measures is the quick identification of the $k$ most central nodes without computing the score of each node~\cite{DBLP:conf/alenex/BergaminiBCMM16,DBLP:conf/www/LeeC14}. 

There are several contexts in which having a high betweenness can be beneficial for the node itself.
For example, in the field of transportation network analysis, the betweenness centrality seems to be positively related to the efficiency of an airport (see~\cite{Malighetti09}, where a network of 57 European airports has been analyzed).
Also, increasing the betweenness of an airport would mean more traffic flowing through it and  possibly more customers for its shops. 
In the context of social networks, Valente and Fujimoto~\cite{DBLP:journals/socnet/ValenteF10} claim that brokers (or ``bridging individuals'') ``may be more effective at changing others, more open to change themselves, and intrinsically interesting to identify''. In~\cite{DBLP:journals/socnet/EverettV16}, the authors show that a slightly modified version of betweenness centrality can be used to find brokers. Also, the authors of~\cite{DBLP:conf/kdd/MahmoodyTU16} show experimentally that nodes with high betweenness are also very effective in spreading influence to other nodes in a social network. Therefore, it might be interesting for a user to create new links with other users or pages in order to increase his own influence spread.


The problem of increasing the centrality of a node has attracted considerable attention for page-rank~\cite{AN06,OV14}, where much effort has been devoted to ``fooling'' search engines in order to increase the popularity of some web pages (an example is the well-known link farming~\cite{DBLP:conf/www/WuD05}).  In addition to page-rank, the problem has been considered also for other centrality measures, such as closeness centrality~\cite{CDSV16} and eccentricity~\cite{DZ10,PBZ13}.

In the above mentioned contexts, it is reasonable to assume that, in order to increase its betweenness, a node can only add edges incident to itself. Hence, in this paper we address the following problem: assuming that a node $v$ can connect itself with $k$ other nodes, how can we choose these nodes in order to maximize the betweenness centrality of $v$? In other terms we want to add a set of $k$ edges to the graph (all incident to $v$), such that the betweenness of $v$ in the new graph is as high as possible. 
For directed graphs, we assume the edges we want to add are of the form $(w, v)$ (i.e. \textit{incoming edges}). However, our results apply also to the problem where $k$ \textit{outgoing edges} need to be added. Indeed, in our proofs, we could simply use $G$ transposed instead of $G$, and the results would also be valid in the case where we want to add outgoing edges.

Since in some contexts one might be more interested in having a high ranking among other nodes rather than a high betweenness score, we also consider the case where we want to maximize the ranking increment of a node instead of its betweenness. We call such two optimization problems \emph{maximum betweenness improvement} (\MBI) and \emph{maximum ranking improvement} (\MRI), respectively.

\paragraph*{Our contribution} 
We study both \MBI and \MRI problems in directed graphs.
Our contribution can be summarized as follows:
$(i)$ We provide two hardness results, one for \MBI and one for \MRI. In particular, we prove that, unless $P = NP$, \MBI cannot be approximated within a factor greater than $1 - \frac{1}{2e}$. Also, we show that, for any constant $\alpha \leq 1$, there is no $\alpha -$approximation algorithm for \MRI, unless $P = NP$ (Section~\ref{sec:hardness}).
$(ii)$ We propose a greedy algorithm for \MBI, which yields a $(1 - \frac{1}{e}) - $approximation (Section~\ref{sec:greedy}). This is in contrast with the results for the undirected graph case, where it is known that the same algorithm has an unbounded approximation ratio~\cite{DSV15}.  The complexity of the algorithm, if implemented naively, is $O(k n^2 m)$. 
$(iii)$ To make our greedy approach faster, we also develop a new algorithm for recomputing the betweenness centrality of a single node after an edge insertion or a weight decrease (Section~\ref{sec:dynamic-single}). The algorithm, which might be of independent interest, builds on a recent method for updating the betweenness of all nodes~\cite{DBLP:conf/sea/BergaminiMOS17}. In the worst case, our algorithm updates the betweenness of one node in $O(n^2)$ time, whereas all existing dynamic algorithms have a worst-case complexity of at least $\Theta(n m)$. This is in contrast with the static case, where computing betweenness of all nodes is just as expensive as computing it for one node (at least, no algorithm exists that computes the betweenness of one node faster than for all nodes). In a context where the betweenness centrality of a single node needs to be recomputed, our experimental evaluation (Section~\ref{sec:experiments}) shows that our new algorithm is much faster than existing algorithms, on average by a factor $18$ for directed and $29$ for undirected graphs (geometric mean of the speedups). Also, using our dynamic algorithm, the worst-case complexity of our greedy approach for \MBI decreases to $O(k n^3)$. However, our experiments show that it is actually much faster in practice. For example, we are able to target directed networks with hundreds of thousands of nodes in a few minutes.

In terms of solution quality, our experiments in Section~\ref{sec:experiments_greedy} show that on directed random graphs, the approximation ratio (the ratio between the solution found by the optimum and the one found by our greedy algorithm) is never smaller than 0.96 for the instances used.
 Also, we show that on real-world networks the greedy approach outperforms other heuristics, both in terms of betweenness improvement and ranking improvement. Although the approximation guarantee holds only for directed graphs, our tests show that the greedy algorithm works well also on undirected real-world networks.


\section{Related work}
\label{sec:related}
\paragraph*{Centrality improvement}
In the following we describe the literature about algorithms that aim at optimizing some property of a graph by adding a limited number of edges.
In~\cite{MT09}, the authors give a constant factor approximation algorithm for the problem of minimizing the average shortest-path distance between all pairs of nodes. Other works~\cite{PBG11,PPP15} propose new algorithms for the same problem and show experimentally that they are good in practice.
In~\cite{BDDSW12}, the authors study the problem of minimizing the average number of hops in shortest paths of weighted graphs, and prove that the problem cannot be approximated within a logarithmic factor, unless $P=NP$. They also propose two approximation algorithms with non-constant approximation guarantees.
\cite{TBFF12} and~\cite{SABA15} focus on the problem of maximizing the leading eigenvalue of the adjacency matrix and give algorithms with proven approximation guarantees.

Some algorithms with proven approximation guarantees for the problem of minimizing the diameter of a graph are presented in~\cite{BGP12} and~\cite{FGGM15}.

In~\cite{LY14} and~\cite{DFHY15}, the authors propose approximation algorithms with proven guarantees for the problem of making the number of triangles in a graph minimum and maximum, respectively.
In~\cite{P15}, the author studies the problem of minimizing the characteristic path length.
 
The problem analyzed in this paper differs from the above mentioned ones as it focuses on improving the centrality of a predefined vertex.
Similar problems have been studied for other centrality measures, i.e. page-rank~\cite{AN06,OV14}, eccentricity~\cite{DZ10,PBZ13}, average distance~\cite{MT09}, some measures related to the number of paths passing through a given node~\cite{IETB12}, and closeness centrality~\cite{CDSV15,CDSV16}. In particular, in~\cite{CDSV16} the authors study the problem of adding a limited amount of edges incident to a target node in order to increase its harmonic centrality (a variant of closeness). They prove that the problem cannot be approximated within a factor greater than $1-\frac{1}{3e}$ ($1-\frac{1}{15e}$ on undirected graphs) and they design a $1-\frac{1}{e}$-approximation algorithm to solve it. They also make use of heuristics to decrease the computational time and  run experiments on large real-world networks.
In this work we show how to adapt the greedy algorithm presented in~\cite{CDSV16} according to the definition of betweenness centrality in order to study the \MBI problem.

The \MBI problem has been studied for undirected weighted graphs~\cite{DSV15} and it has been proved that, in this case, the problem cannot be approximated within a factor greater than $1-\frac{1}{2e}$, unless $P=NP$. They proved this bound using a technique similar to the one used in~\cite{CDSV15} for the harmonic centrality (and to the one used in this paper for directed graphs).
Also, {D'Angelo} et al.~\cite{DSV15} show that a natural greedy algorithm exhibits an arbitrarily small approximation ratio. Nevertheless, in their experiments on small networks with up to few hundreds of nodes, they show that the greedy algorithm provides a solution near to the optimal.
In this paper, we make the greedy algorithm orders of magnitude faster by combining it with a new dynamic algorithm for updating the betweenness of one node and we study the behavior of the algorithm on directed and undirected networks with up to $10^4$ nodes and $10^5$ edges.


\paragraph*{Dynamic algorithms for betweenness centrality}
The general idea of dynamic betweenness algorithms is to keep track of the old betweenness values and to update them after some modification happens to the graph, which might be an edge or node insertion, an edge or node deletion, or a change in an edge's weight. In particular, in case of edge insertions or weight decreases, the algorithms are often referred to as \textit{incremental}, whereas for edge deletions or weight increases they are called \textit{decremental}.
All dynamic algorithms existing in the literature update the centralities of \textit{all nodes} and most of them first update the distances and shortest paths between nodes and then recompute the fraction of shortest paths each node belongs to. The approach proposed
by Green \etal~\cite{DBLP:conf/socialcom/GreenMB12} for unweighted graphs 
maintains all previously calculated betweenness values and
additional information, like the distance
between each node pair and the list of \textit{predecessors}, i.e.\ the nodes
immediately preceding $v$ in the shortest paths from $s$ to $v$, for all node pairs $(s,v)$. Using this information,
the algorithm limits the recomputation to the nodes whose
betweenness has actually been affected. Kourtellis
\etal~\cite{kourtellis2014scalable} modify the
approach by Green \etal~\cite{DBLP:conf/socialcom/GreenMB12} in
order to reduce the memory requirements from $O(nm)$ to $O(n^2)$. Instead of
storing the predecessors of each node $v$ from each possible
source, they recompute them every time the information is required. 

Kas \etal~\cite{DBLP:conf/asunam/KasWCC13} extend an existing algorithm for
the dynamic all-pairs shortest paths (APSP) problem by Ramalingam and Reps~\cite{DBLP:journals/tcs/RamalingamR96}
to also update BC scores. 
Nasre \etal~\cite{DBLP:conf/mfcs/NasrePR14} compare the distances between each node pair before and after the update and then recompute the dependencies as in Brandes's algorithm. Although this algorithm is faster than recomputation on some graph classes (i.e.\ when only edge insertions are allowed and the graph is sparse and weighted), it was shown in~\cite{DBLP:conf/alenex/BergaminiMS15} that its performance in practice is always worse than that of the algorithm proposed in~\cite{DBLP:conf/socialcom/GreenMB12}. Pontecorvi and Ramachandran~\cite{DBLP:conf/isaac/PontecorviR15} extend existing fully-dynamic APSP algorithms with new data structures to update \textit{all} shortest paths and then recompute dependencies as in Brandes's algorithm. 
Differently from the previous algorithms, the approach by Lee \etal~\cite{DBLP:journals/isci/LeeCC16} is not based on dynamic APSP algorithms, but splits the graph into biconnected components and then recomputes the betweenness values from scratch only within the component affected by the graph update. Although this allows for a smaller memory requirement ($\Theta(m)$ versus $\Omega(n^2)$ needed by the other approaches), the speedups on recomputation reported in~\cite{DBLP:journals/isci/LeeCC16} are significantly worse than those reported for example in~\cite{DBLP:conf/socialcom/GreenMB12} or Kas \etal~\cite{DBLP:conf/asunam/KasWCC13}.

Very recently, a new approach called \textsf{iBet} for updating betweenness after an edge insertion or a weight decrease has been proposed~\cite{DBLP:conf/sea/BergaminiMOS17}. The approach improves over the one by Kas \etal~\cite{DBLP:conf/asunam/KasWCC13} by removing redundant work in both the APSP update step and the dependency accumulation. In their experiments, the authors show that \textsf{iBet} outperforms existing dynamic algorithms by about one order of magnitude. Since our new dynamic algorithm for updating the betweenness of a single node builds on \textsf{iBet}, we will describe it in more detail in Section~\ref{sec:quinca}.

Recently, also dynamic algorithms that update an approximation of betweenness centrality have been proposed~\cite{DBLP:journals/corr/BergaminiM16,DBLP:journals/pvldb/HayashiAY15,DBLP:journals/corr/RiondatoU16}.
Notice that all existing dynamic algorithms update the betweenness of all nodes and their worst-case complexity is, in general, the same as static recomputation. This means, for exact algorithms, $O(n m)$ in unweighted and $O(n(m+n \log n))$ in weighted graphs.

\section{Notation and problem statement}
\label{sec:notation}

Let $G=(V,E)$ be a directed graph where $|V|=n$ and $|E|=m$. For each node $v$, $N_v$ denotes the set of in-neighbors of $v$, i.e. $N_v=\{u~|~(u,v)\in E\}$.
Given two nodes $s$ and $t$, we denote by $d_{st}$, $\sigma_{st}$, and $\sigma_{stv}$ the distance from $s$ to $t$ in $G$, the number of shortest paths from $s$ to $t$ in $G$, and the number of shortest paths from $s$ to $t$ in $G$ that contain $v$, respectively. For each node pair $(s, t$), we assume $d_{st} \geq 0$. For each node $v$, the \emph{betweenness centrality}~\cite{F77} of $v$ is defined as
\begin{equation}
\label{def:bc}
 b_v = \sum_{\substack{s,t\in V\\s\neq t;s,t\neq v\\\sigma_{st}\neq 0}}\frac{\sigma_{stv}}{\sigma_{st}}.
\end{equation}
In case $\sigma_{st} = 0$, the corresponding term in the sum is defined to be 0.  The \emph{ranking} of each node $v$ according to its betweenness centrality is defined as
\begin{equation}
\label{def:rank}
 r_v = |\{u\in V~|~b_u>b_v\}| + 1.
\end{equation}

In this paper, we consider graphs that are augmented by adding a set $S$ of arcs not in $E$. Given a set $S\subseteq V\times V\setminus E$ of arcs, we denote by $G(S)$ the graph augmented by adding the arcs in $S$ to $G$, i.e. $G(S)=(V,E\cup S)$. For a parameter $x$ of $G$, we denote by $x(S)$ the same parameter in graph $G(S)$, e.g. the distance from $s$ to $t$ in $G(S)$ is denoted as $d_{st}(S)$.

The betweenness centrality of a node might change if the graph is augmented with a set of arcs. In particular, adding arcs incident to some node $v$ can increase the betweenness of $v$ and its ranking. We are interested in finding a set $S$ of arcs incident to a particular node $v$ that maximizes $b_v(S)$. Therefore, we define the following optimization problem.
\optproblem{Maximum Betweenness Improvement (\MBI)}
{A directed graph $G=(V,E)$; a node $v\in V$; an integer $k\in\mathbb{N}$}
{A set $S$ of arcs incident to $v$, $S=\{(u,v)~|~u\in V\setminus N_v\}$, such that $|S|\leq k$}
{Maximize $b_v(S)$}

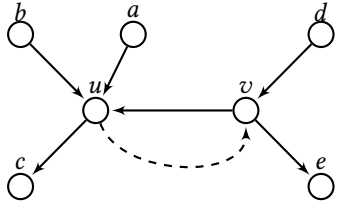
\begin{figure}[t]
 	\centering
   \begin{tikzpicture}[> = latex',shorten >=1pt,auto,node distance=3cm,
   thick,main node/.style={circle,draw}]  
    \node (a) at (7.5,3.65) {$a$};
    \node[main node] (a) at (7.5,3.35) {};

      \node (b) at (6,3.65) {$b$};
      \node[main node] (b) at (6,3.35) {};

      \node (c) at (6,1.65) {$c$};
      \node[main node] (c) at (6,1.35) {};
   
      \node (u) at (7,2.65) {$u$};
      \node[main node] (u) at (7,2.35) {};
   \node (v) at (9,2.65) {$v$};
   \node[main node] (v) at (9,2.35) {};

   \node (d) at (10,3.65) {$d$};
   \node[main node] (d) at (10,3.35) {};

   \node (e) at (10,1.65) {$e$};
   \node[main node] (e) at (10,1.35) {};

     \draw[->, dashed] (u) to[out=-70,in=-90] (v);
     \draw[->] (u) to (c);
     \draw[->] (v) to (u);
   \draw[->] (b) to (u);
   \draw[->] (a) to (u);
      \draw[->] (d) to (v);
      \draw[->] (v) to (e);

   \end{tikzpicture}
 	\caption{Graph in which the addition of the edge $(u,v)$ affects the betweenness value but not the ranking.}
 	\label{fig:example}
 \end{figure}

Note that maximizing the betweenness value of a node $v$ does not necessarily lead to maximizing the ranking position of $v$.
For example, consider the graph in Figure~\ref{fig:example}: before the addition of the edge $(u,v)$ the initial betweenness values are $b_u =2$, $b_v =1$ and $b_a = b_b = b_c = b_d = b_e = 0$ while the initial ranking is $r_u=1$, $r_v=2$ and $r_a = r_b = r_c = r_d = r_e = 0$. After the addition of the edge $(u,v)$ the new betweenness values are $b'_u =6$, $b'_v =4$ and $b_a = b_b = b_c = b_d = b_e = 0$ but the ranking remains the same.  

Therefore, we also consider the problem of finding a set $S$ of arcs incident to node $v$ that maximizes \emph{the increment of the ranking} of $v$ with respect to its original ranking. We denote such an increment as $\rho_v(S)$, that is,
\[
 \rho_v(S) = r_v - r_v(S).
\]
Informally, $\rho(S)$ represents the number of nodes that $v$ ``overtakes'' by adding arcs in $S$ to $G$. Therefore, we define the following optimization problem.
\optproblem{Maximum Ranking Improvement (\MRI)}
{A directed graph $G=(V,E)$; a vertex $v\in V$; and an integer $k\in\mathbb{N}$}
{A set $S$ of arcs incident to $v$, $S=\{(u,v)~|~u\in V\setminus N(v)\}$, such that $|S|\leq k$}
{Maximize $\rho_v(S)$}

\section{Hardness of approximation}
\label{sec:hardness}
In this section we first show that it is $NP$-hard to approximate problem \MBI within a factor greater than $1-\frac{1}{2e}$. Then, we focus on the $\MRI$ problem and show that it cannot be approximated within any constant bound, unless $P=NP$.

\begin{theorem}\label{th:inapx:bet}
 Problem \MBI cannot be approximated within a factor greater than $1-\frac{1}{2e}$, unless $P=NP$.
\end{theorem}
\begin{proof}
We give an L-reduction with parameters $a$ and $b$~\cite[Chapter~16]{WS11} to the \emph{maximum set coverage problem} (\MSC) defined as follows: given a finite set $X$, a finite family $\mathcal{F}$ of subsets of $X$, and an integer $k'$, find $\mathcal{F}'\subseteq\mathcal{F}$ such that $|\mathcal{F}'|\leq k'$ and $s(\mathcal{F}') = |\cup_{S_i\in \mathcal{F}'}S_i|$ is maximum.
In detail, we will give a polynomial-time algorithm that transforms any instance $I_\MSC$ of \MSC into an instance $I_\MBI$ of $\MBI$ and a polynomial-time algorithm that transforms any solution $S_\MBI$ for $I_\MBI$ into a solution $S_\MSC$ for $I_\MSC$ such that the following two conditions are satisfied for some values $a$ and $b$:
 \begin{align}
  OPT(I_\MBI)&\leq a OPT(I_\MSC)\label{lreduction:bet:one}\\
  OPT(I_\MSC) - s(S_\MSC) &\leq b \left(OPT(I_\MBI) - b_v(S_\MBI)\right),\label{lreduction:bet:two}
 \end{align}
 where $OPT$ denotes the optimal value of an instance of an optimization problem. 
If the above conditions are satisfied and there exists an $\alpha$-approximation algorithm $A_\MBI$ for \MBI, then there exists a $(1-ab(1-\alpha))$-approximation algorithm $A_\MSC$ for \MSC~\cite[Chapter 16]{WS11}. 
Since it is $NP$-hard to approximate \MSC within a factor greater than $1-\frac{1}{e}$~\cite{F98}, then the approximation factor of $A_\MSC$ must be smaller than $1-\frac{1}{e}$, unless $P=NP$. This implies that $1-ab(1-\alpha) < 1-\frac{1}{e}$ that is, the approximation factor $\alpha$ of $A_\MBI$ must satisfy $\alpha < 1-\frac{1}{abe}$, unless $P=NP$. In the following, we give an L-reduction and determine the constant parameters $a$ and $b$. In the reduction, each element $x_i$ and each set $S_j$ in an instance of \MSC corresponds to a vertex in an instance of \MBI, denoted by $v_{x_i}$ and $v_{S_j}$, respectively. There is an arc from $v_{x_i}$ to $v_{S_j}$ if and only if $x_i\in S_j$. The \MBI instance contains two further nodes $v$ and $t$ and an arc $(v,t)$. A solution to such an instance consists of arcs from nodes $v_{S_j}$ to $v$ and the aim is to cover with such arcs the maximum number of shortest paths from nodes $v_{x_i}$ to $t$. We will prove that we can transform a solution to \MBI into a solution to \MSC such that any node $v_{x_i}$ that has a shortest path passing trough $v$ corresponds to a covered element $x_i\in X$. We give more detail in what follows.

 \begin{figure}[t]
 \centering
  \begin{tikzpicture}[->,>=stealth',shorten >=1pt,auto,node distance=3cm,
  thick,main node/.style={circle,draw}]
  \tikzset{arc/.style = {->,> = latex'}}
  \node (x1) at (0,3.3) {$v_{x_1}$};
  \node[main node] (x11) at (0,3) {};
  
  \node (x2) at (0,2.3) {$v_{x_2}$};
  \node[main node] (x21) at (0,2) {};

  \node (x3) at (0,1.2) {$\vdots$};

  \node (xn) at (0,0.3) {$v_{x_{|X|}}$};
  \node[main node] (xn1) at (0,0) {};

  \node (s1) at (2.5,3.3) {$v_{S_1}$};
  \node[main node] (s11) at (2.5,3) {};
  
  \node (s2) at (2.5,2.3) {$v_{S_2}$};
  \node[main node] (s21) at (2.5,2) {};

  \node (s3) at (2.5,1.2) {$\vdots$};

  \node (sn) at (2.5,0.3) {$v_{S_{|\mathcal{F}|}}$};
  \node[main node] (sn1) at (2.5,0) {};

  \node (v) at (6,2) {$v$};
  \node[main node] (v1) at (6,1.7) {};
  \node (t) at (8,2) {$t$};
  \node[main node] (t1) at (8,1.7) {};  
  
  \draw[arc] (v1) to (t1);
  \draw[arc] (x11) to (s21);
  \draw[arc] (x11) to (s11);  
  \draw[arc] (x21) to (s11);    
  \draw[arc] (x21) to (sn1);   
  \draw[->, dashed] (s11) to (v1);
  \draw[->, dashed] (s21) to (v1);
  \draw[->, dashed] (sn1) to (v1);
\end{tikzpicture}
\caption{Reduction used in Theorem~\ref{th:inapx:bet}. In the example $x_1\in S_1$, $x_1\in S_2$, $x_2\in S_1$, and $x_2\in S_{\mathcal{F}}$. The dashed arcs denote those added in a solution.}
\label{fig:inapx:bet}
 \end{figure}
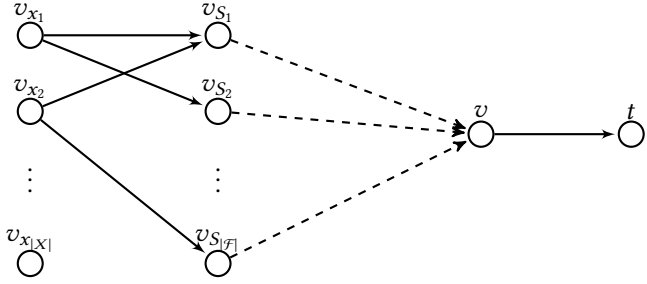

 Given an instance $I_\MSC=(X,\mathcal{F},k')$ of \MSC, where $\mathcal{F} = \{S_1,S_2,\ldots S_{|\mathcal{F}|}\}$, we define an instance $I_\MBI=(G,v,k)$ of \MBI, where:
 \begin{itemize}
  \item $G=(V,E)$;
  \item $V = \{ v,t \} \cup \{v_{x_i}~|~x_i\in X\} \cup \{v_{S_j}~|~S_j\in \mathcal{F}\}$;
  \item $E = \{(v,t)\} \cup \{(v_{x_i},v_{S_j})~|~x_i\in S_j\}$;
  \item $k=k'$.
 \end{itemize}
 See Figure~\ref{fig:inapx:bet} for a visualization.
 
 Without loss of generality, we can assume that any solution $S_\MBI$ to \MBI contains only arcs $(v_{S_j},v)$ for some $S_j\in \mathcal{F}$. In fact, if a solution does not satisfy this property, then we can improve it in polynomial time by repeatedly applying the following transformation: for each arc $a = (v_{x_i},v)$ in $S_\MBI$ such that $x_i\in X$, exchange $a$ with an arc $(v_{S_j},v)$ such that $x_i\in S_j$ and $(v_{S_j},v)\not\in S_\MBI$ if it exists or remove $a$ otherwise. Note that if no arc $(v_{S_j},v)$ such that $x_i\in S_j$ and $(v_{S_j},v)\not\in S_\MBI$ exists, then all the shortest paths from $x_i$ to $t$ pass through $v$ and therefore the arc $(v_{x_i},v)$ can be removed without changing the value of $b_v(S_\MBI)$.
 Such a transformation does not decrease the value of $b_v(S_\MBI)$ in fact, all the shortest paths passing through $v$ in the original solution still pass through $v$ in the obtained solution. Moreover, if Condition~(\ref{lreduction:bet:two}) is satisfied for the obtained solution, then it is satisfied also for the original solution.
 In such a solution, all the paths (if any) from $v_{x_i}$ to $t$, for each $x_i\in X$, and from $v_{S_j}$ to $t$, for each $S_j\in \mathcal{F}$ pass through $v$ and therefore the ratio $\frac{\sigma_{stv}(S_\MBI)}{\sigma_{st}(S_\MBI)}$ is 1, for each $s\in V\setminus \{v,t\}$ such that $\sigma_{st}(S_\MBI)\neq 0$. We can further assume, again without loss of generality, that any solution $S_\MBI$ is such that $|S_\MBI|=k$, in fact, if $|S_\MBI|<k$, then we can add  to $S_\MBI$ an arc $(v_{S_j},v)$ that is not yet in $S_\MBI$. Note that such an arc must exists otherwise $k>|\mathcal{F}|$ and this operation does not decrease the value of $b_v(S_\MBI)$.
 
 Given a solution $S_\MBI=\{(v_{S_j},v)~|~S_j\in\mathcal{F}\}$ to \MBI, we construct the solution $S_\MSC=\{S_j~|~(v_{S_j},v)\in S_\MBI\}$ to \MSC.  By construction, $|S_\MSC| = |S_\MBI| = k = k'$. Moreover, the set of elements $x_i$ of $X$  such that $\sigma_{v_{x_i}t}(S_\MBI)\neq 0$ is equal to $\{x_i\in S_j~|~(v_{S_j},v)\in S_\MBI\} =\bigcup_{S_j\in S_\MSC}S_j$.
 Therefore, the betweenness centrality of $v$ in $G(S_\MBI)$ is:
\begin{align*}
   b_v(S_\MBI) =& \sum_{\substack{s\in V\setminus \{v,t\}\\\sigma_{st}(S_\MBI)\neq 0}}\frac{\sigma_{stv}(S_\MBI)}{\sigma_{st}(S_\MBI)}\\
  =&\sum_{\substack{x_i\in X\\\sigma_{v_{x_i}t}(S_\MBI)\neq 0}}\frac{\sigma_{v_{x_i}tv}(S_\MBI)}{\sigma_{v_{x_i}t}(S_\MBI)} + \sum_{\substack{S_j\in \mathcal{F}\\\sigma_{v_{S_j}t}(S_\MBI)\neq 0}}\frac{\sigma_{v_{S_j}tv}(S_\MBI)}{\sigma_{v_{S_j}t}(S_\MBI)}\\
  =&|\{x_i\in S_j~|~(v_{S_j},v)\in S_\MBI\}| + |\{S_j~|~ (v_{S_j},v)\in S_\MBI\}|\\
  =&\left|\bigcup_{S_j\in S_\MSC}S_j\right| + |S_\MSC|\\
  =&s(S_\MSC) + k.
\end{align*}
It follows that Conditions~(\ref{lreduction:bet:one}) and~(\ref{lreduction:bet:two}) are satisfied for $a=2$, $b=1$ since: $OPT(I_\MBI) =  OPT(I_\MSC) + k \leq 2OPT(I_\MSC)$ and $OPT(I_\MSC) - s(S_\MSC) = OPT(I_\MBI) - b_v(S_\MBI)$, where the first inequality is due to the fact that $OPT(I_\MSC)\geq k$.\footnote{If $OPT(I_\MSC) < k$, then the greedy algorithm finds an optimal solution for \MSC.}
The statement follows by plugging the values of $a$ and $b$ into $\alpha < 1-\frac{1}{abe}$.\qed
\end{proof}

In the next theorem, we show that, unless $P=NP$, we cannot find a polynomial time approximation algorithm for \MRI with a constant approximation guarantee. 

\begin{theorem}\label{th:inapx:ranking}
For any constant $\alpha\leq 1$, there is no $\alpha$-approximation algorithm for the \MRI problem, unless $P=NP$.
\end{theorem}
\begin{proof}
 By contradiction, let us assume that there exists a polynomial time algorithm $A$ that guarantees an approximation factor of $\alpha$. We show that we can use $A$ to determine whether an instance $I$ of the \emph{exact cover by 3-sets} problem (\XC) admits a feasible solution or not. Problem \XC is known to be $NP$-complete~\cite{GJ79} and therefore this implies a contradiction. In the \XC problem we are given a finite set $X$ with $|X| = 3q$ and a collection $C$ of 3-element subsets of $X$ and we ask whether $C$ contains an exact cover for $X$, that is, a subcollection $C' \subseteq C$ such that every element of $X$ occurs in exactly one member of $C'$. Note that we can assume without loss of generality that $m>q$.
 
  \begin{figure}[t]
  \centering
  \scalebox{0.8}{  \begin{tikzpicture}[> = latex',shorten >=1pt,auto,node distance=3cm,
  thick,main node/.style={circle,draw}]
  \node (x1) at (0,5.7) {$v_{x_1}$};
  \node[main node] (x11) at (0,5.4) {};
  
  \node (x2) at (0,3.7) {$v_{x_2}$};
  \node[main node] (x21) at (0,3.4) {};

  \node (x3) at (0,1.7) {$\vdots$};

  \node (xn) at (0,0.3) {$v_{x_n}$};
  \node[main node] (xn1) at (0,0) {};

  \node (s1) at (2.5,5.7) {$v_{T_1}$};
  \node[main node] (s11) at (2.5,5.4) {};
  
  \node (s2) at (2.5,3.7) {$v_{T_2}$};
  \node[main node] (s21) at (2.5,3.4) {};

  \node (s3) at (2.5,1.7) {$\vdots$};

  \node (sn) at (2.5,0.3) {$v_{T_m}$};
  \node[main node] (sn1) at (2.5,0) {};

  \node (m11) at (5,7.3) {$v_{T_1^1}$};
  \node[main node] (m111) at (5,7) {};
  
  \node (m12) at (5,6.6) {$v_{T_1^2}$};
  \node[main node] (m121) at (5,6.3) {};

  \node (m13) at (5,5.8) {$\vdots$};
  
  \node (m14) at (5,5.2) {$v_{T_1^M}$};
  \node[main node] (m141) at (5,4.9) {};

  \node (m21) at (5,4.3) {$v_{T_2^1}$};
  \node[main node] (m211) at (5,4) {};
  
  \node (m22) at (5,3.6) {$v_{T_2^2}$};
  \node[main node] (m221) at (5,3.3) {};

  \node (m23) at (5,2.8) {$\vdots$};
  
  \node (m24) at (5,2.2) {$v_{T_2^M}$};
  \node[main node] (m241) at (5,1.9) {};

  \node (m31) at (5,1.3) {$v_{T_m^1}$};
  \node[main node] (m311) at (5,1) {};
  
  \node (m32) at (5,0.6) {$v_{T_m^2}$};
  \node[main node] (m321) at (5,0.3) {};

  \node (m33) at (5,-0.2) {$\vdots$};
  
  \node (m34) at (5,-0.8) {$v_{T_m^M}$};
  \node[main node] (m341) at (5,-1.1) {};

  \node (v) at (8.5,2.65) {$v$};
  \node[main node] (v1) at (8.5,2.35) {};

  \node (u) at (10,3.65) {$u$};
  \node[main node] (u1) at (10,3.35) {};

  \node (t1) at (10,1.65) {$t_1$};
  \node[main node] (t11) at (10,1.35) {};
  \node (t2) at (10,0.65) {$t_2$};
  \node[main node] (t21) at (10,0.35) {};
  \node (t3) at (10,-0.35) {$t_3$};
  \node[main node] (t31) at (10,-0.65) {};

  \draw[->] (x11) to (s21);
  \draw[->] (x11) to (s11);  
  \draw[->] (x21) to (s11);    
  \draw[->] (x21) to (sn1);

  \draw[->] (s11) to (m111);
  \draw[->] (s11) to (m121);
  \draw[->] (s11) to (m141);

  \draw[->] (s21) to (m211);
  \draw[->] (s21) to (m221);
  \draw[->] (s21) to (m241);  

  \draw[->] (sn1) to (m311);
  \draw[->] (sn1) to (m321);
  \draw[->] (sn1) to (m341);

  \draw[->, dashed] (m111) to (v1);
  \draw[->, dashed] (m211) to (v1);
  \draw[->, dashed] (m311) to (v1);
  
  \draw[->] (u1) to (v1);
  \draw[->] (v1) to (t11);
  \draw[->] (v1) to (t21);
  \draw[->] (v1) to (t31);
\end{tikzpicture}}
  \caption{The reduction used in Theorem~\ref{th:inapx:ranking}. The dashed arcs denote those added in a solution to \MRI.}
  \label{fig:inapx:ranking}
  \end{figure}
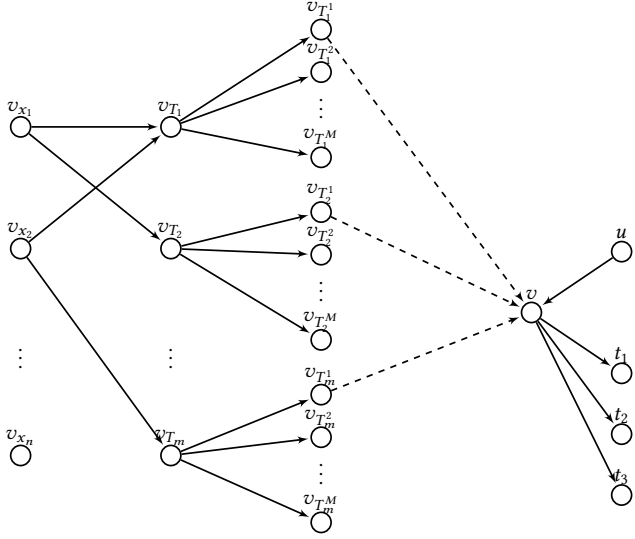
 
 Given an instance $I=(X,C)$ of \XC where $|X|=n=3q$ and $|C|=m$, we define an instance $I'=(G,v,k)$ of \MRI as follows.
 \begin{itemize}
  \item $G=(V,E)$;
  \item $V= \{v,u,t_1,t_2,t_3\} \cup \{ v_{x_i}~|~x_i\in X\} \cup \{v_{T_j}~|~T_j\in C\} \cup \{ v_{T_j^\ell}~|~T_j\in C,~ \ell = 1,2,\ldots,M\}$;
  \item $E= \{(v_{x_i},v_{T_j})~|~x_i\in T_j\} \cup \{ (v_{T_j},v_{T_j^\ell})~|~T_j\in C,~ \ell = 1,2,\ldots,M\} \cup \{(u,v),(v,t_1),(v,t_2),(v,t_3)\}$;
  \item $k=q$.
 \end{itemize}
 where $M=5q+1$. See Figure~\ref{fig:inapx:ranking} for a visualization.
   
 The proof proceeds by showing that $I$ admits an exact cover if and only if $I'$ admits a solution $S$ such that $\rho_v(S) > 0$. This implies that, if $OPT$ is an optimal solution for $I'$, then $\rho_v(OPT) > 0 $ if and only if $I$ admits an exact cover. Hence, the statement follows by observing that algorithm $A$ outputs a solution $S$ such that $\rho_v(S) > \alpha\rho_v(OPT)$ and hence $\rho_v(S) > 0$ if and only if $I$ admits an exact cover.
 
 In $I'$, $b_v = 3$, $b_{v_{T_j}} = 3M = 15q + 3$, for each $T_j\in C$, and $b_w=0$, for any other node $w$. Therefore, $r_{T_j} = 1$, for each $T_j\in C$, $r_v=m+1$, and $r_w=m+2$, for any other node $w$. In the proof we will use the observation that, in instance $I'$, adding arcs incident to $v$ does not decrease the betweenness value of any node, that is for any node $w\in V$ and for any solution $S$ to $I'$, $b_w(S)\geq b_w$.
 
 If instance $I$ of \XC admits an exact cover $C'$, then consider the solution $S = \{(v_{T_{j}^1},v) ~|~ T_{j}\in C'\}$ to $I'$. Note that $|S| = q = k$ and therefore we only need to show that $\rho_v(S)> 0$. Indeed, in the following we show that $\rho_v(S) = m-q>0$. Since $C'$ is an exact cover, then all  nodes $v_{x_i}$ are connected to the 3 nodes $t_i$ and all the paths connecting them pass through $v$. The same holds for nodes $v_{T_{j}}$ and $v_{T_{j}^1}$ such that $T_{j}\in C'$. Since there are $3q$ nodes $v_{x_i}$, $q$ nodes $v_{T_{j}}$ such that $T_{j}\in C'$, and $q$ nodes $v_{T_{j}^1}$ such that $T_{j}\in C'$, then the betweenness centrality of $v$ increases to $b_v(S) = 3(5q+1) = 15q + 3$. Nodes  $v_{T_{j}}$ and $v_{T_{j}^1}$ such that $T_{j}\in C'$ increase their centrality to $b_{v_{T_{j}}}(S)=3(M+4) = 15q+15$ and  $b_{v_{T_{j}^1}}(S)=16$, respectively. Any other node does not change its betweenness centrality. Therefore the only nodes that have a betweenness higher than $v$ are the $q$ nodes $v_{T_{j}^1}$ such that $T_{j}\in C'$. It follows that $r_v(S) = q+1$ and $\rho_v(S) = m+1 - (q+1) = m-q>0$.

 Let us now assume that $I'$ admits solution $S$ such that $|S|\leq k$ and $\rho_v(S)>0$. We first prove that $S$ is only made of arcs in the form $(v_{T_{j}^1},v)$ and that $b_v(S)\geq 15q+3$ or that it can be transformed in polynomial time into a solution with such a form without increasing its size. Assume that $S$ has arcs not in this form, then we can apply one of the following transformations to each of such arcs $e=(w,v)$.
 \begin{itemize}
  \item If $w=v_{x_i}$ for some $x_i\in X$ and there exists a node $v_{T_{j}^1}$ such that $x_i\in T_{j}$ and $(v_{T_{j}^1},v)\not\in S$, then remove $e$ and add arc $(v_{T_{j}^1},v)$ to $S$;
  \item If $w=v_{x_i}$ for some $x_i\in X$ and $(v_{T_{j}^1},v)\in S$ for all $T_{j}$ such that $x_i\in T_{j}$, then remove $e$;
  \item If $w=v_{T_{j}}$ for some $T_j\in C$ and $(v_{T_{j}^1},v)\not\in S$, then remove $e$ and add arc $(v_{T_{j}^1},v)$ to $S$;
  \item If $w=v_{T_{j}}$ for some $T_j\in C$ and $(v_{T_{j}^1},v)\in S$, then remove $e$;
  \item If $w=v_{T_{j}^i}$ for some $T_j\in C$ and $i>1$, and $(v_{T_{j}^1},v)\not\in S$, then remove $e$ and add arc $(v_{T_{j}^1},v)$ to $S$;
  \item If $w=v_{T_{j}^i}$ for some $T_j\in C$ and $i>1$, and $(v_{T_{j}^1},v)\in S$, then remove $e$ and add arc $(v_{T_{j'}^1},v)$ to $S$ for some $j'$ such that $(v_{T_{j'}^1},v)\not \in S$;\footnote{\label{note}Note that such $j'$ must exists, otherwise $m<q$.}
  \item If $w=t_i$ for $i\in\{1,2,3\}$, then remove $e$ and add arc $(v_{T_{j'}^1},v)$ to $S$ for some $j'$ such that $(v_{T_{j'}^1},v)\not \in S$.\footnoteref{note}
 \end{itemize}
 Let us denote by $S'$ and $S$ the original solution and the solution that is eventually obtained by applying the above transformations, respectively. All the above transformations remove an arc and possibly add another arc, therefore the size of the transformed solution is at most the original size, that is $|S| \leq |S'|\leq k$. It remains to show that $\rho_v(S') >0$ implies $b_v(S)\geq 15q+3$. Indeed, observe that $v$ is initially in position $m+1$ and the only nodes that have a betweenness value higher than $v$ are the $m$ nodes $v_{T_{j}}$. Therefore, since $\rho_v(S')>0$, there is at least a node $v_{T_{j}}$ such that $b_v(S') \geq b_{v_{T_{j}}}(S')$. Moreover, all the transformations do not decrease the value of $b_v$ and then $b_v(S)\geq b_v(S')$ and, considering that $b_{v_{T_{j}}}(S') \geq b_{v_{T_{j}}} = 15q+3$, we obtain $b_v(S)\geq 15q+3$. 

 We now prove that the solution $C' = \{T_j~|~(v_{T_{j}^1},v) \in S\}$ to $I$ is an exact cover. By contradiction, let us assume that an element in $X$ is not contained in any set in $C'$ or that an element in $X$ is contained in more than one set in $C'$. The latter case implies the former one since  $|C'| = q$, all the sets in $C'$ contain exactly 3 elements, and $|X|=3q$. Hence, we assume that an element in $|X|$ is not contained in any set in $C'$. This implies that there exists a node $v_{x_i}\in V$ that has no path to nodes $t_i$ and therefore the betweenness of $v$ is at most $3(1+3q-1+2q) =15q$, which is a contradiction to $b_v(S)\geq 15q+3$.\qed
\end{proof}

\section{Greedy approximation algorithm for \MBI}
\label{sec:greedy}
\begin{algorithm2e}[t]
\caption{\Greedy algorithm.}
\label{alg:greedy}
\SetKwInput{Proc}{Algorithm}
\SetKwInOut{Input}{Input}
\SetKwInOut{Output}{Output}
\Input{A directed graph $G=(V,E)$; a vertex $v\in V$; and an integer $k\in\mathbb{N}$}
\Output{Set of edges $S\subseteq \{(u,v)~|~u\in V\setminus N_v\}$ such that $|S|\leq k$}
$S\leftarrow\emptyset$\;
\For{$i=1,2,\ldots,k$} 
{\label{greedy1}
\ForEach{$u\in V\setminus( N_v(S))$}
{ \label{greedy2}
Compute $b_v(S\cup\{(u,v)\})$\; \label{greedy3}
} 
$u_{\max} \leftarrow \arg\max\{b_v(S\cup\{(u,v)\}) ~|~u\in V\setminus( N_v(S))\}$\;
$S \leftarrow S \cup \{(u_{\max},v)\}$\;
}
\Return $S$\;
\end{algorithm2e}

In this section we propose an algorithm that guarantees a constant approximation ratio for the \MBI problem. The algorithm exploits the results of Nemhauser et al. on the approximation of monotone submodular objective functions~\cite{NWF78}. Let us consider the following optimization problem: given a finite set $N$, an integer $k'$, and a real-valued function $z$ defined on the set of subsets of $N$, find a set $S\subseteq N$ such that $|S|\leq k'$ and $z(S)$ is maximum. If $z$ is \emph{monotone and submodular}\footnote{For a ground set $N$, a function $z:2^N\rightarrow \mathbb{R}$ is submodular if for any pair of sets $S\subseteq T \subseteq N$ and for any element $e\in N\setminus T$, $z(S\cup\{e\}) - z(S) \geq z(T\cup \{e\}) - z(T)$.}, then the following greedy algorithm exhibits an approximation of $1-\frac{1}{e}$~\cite{NWF78}: start with the empty set and repeatedly add an element that gives the maximal marginal gain, that is, if $S$ is a partial solution, choose the element $j\in N\setminus S$ that maximizes $z(S\cup\{j\})$.
\begin{theorem}[\cite{NWF78}]\label{thm:submodular}
For a non-negative, monotone submodular function $z$, let $S$ be a set of size $k$ obtained by selecting elements one at a time, each time choosing an element that provides the largest marginal increase in the value of $z$. Then $S$ provides a $\left(1-\frac{1}{e}\right)$-approximation.
\end{theorem}
In this paper we exploit such results by showing that  $b_v$ is monotone and submodular with respect to the possible set of arcs incident to $v$. Hence, we define a greedy algorithm, reported in Algorithm~\ref{alg:greedy}, that provides a  $\left(1-\frac{1}{e}\right)$-approximation. Algorithm~\ref{alg:greedy} iterates $k$ times and, at each iteration, it adds to a solution $S$ an arc $(u,v)$ that, when added to $G(S)$, gives the largest marginal increase in the betweenness of $v$, that is, $b_v(S\cup \{(u,v)\})$ is maximum among all the possible arcs not in $E\cup S$ incident to $v$.
The next theorem shows that the objective function is monotone and submodular.

\begin{theorem}\label{th:subm:bet}
\label{theo:submodular}
 For each node $v$, function $b_v$ is monotone and submodular with respect to any feasible solution for \MBI.
\end{theorem}
\begin{proof}
We prove that each term of the sum in the formula of $b_v$ is monotone increasing and submodular. For each pair $s,t\in V$ such that $s\neq t$ and $s,t\neq v$, we denote such term by $b_{stv}(X) = \frac{\sigma_{stv}(X)}{\sigma_{st}(X)}$, for each solution $X$ to \MBI.

We first give two observations that will be used in the proof. Let $X,Y$ be two solutions to \MBI such that $X\subseteq Y$.
\begin{itemize}
 \item Any shortest path from $s$ to $t$ in $G(X)$ exists also in $G(Y)$. It follows that $d_{st}(Y)\leq d_{st}(X)$.
 
 \item If $d_{st}(Y) < d_{st}(X)$, then any shortest path from $s$ to $t$ in $G(Y)$ passes through arcs in $Y\setminus X$. Therefore, all such paths pass through $v$. It follows that if $d_{st}(Y) < d_{st}(X)$, then $b_{stv}(Y) = 1$. 
\end{itemize}

We now show that $b_v$ is monotone increasing, that is for each solution $S$ to \MBI and for each  node $u$ such that $(u,v)\not\in S\cup E$,
\begin{align*}
 b_{stv}(S\cup\{(u,v)\}) \geq b_{stv}(S).
\end{align*}
If $d_{st}(S)>d_{st}(S\cup\{(u,v)\})$, then $b_{stv}(S\cup\{(u,v)\})=1$ and since by definition $b_{stv}(S)\leq 1$, then the statement holds. If $d_{st}(S)=d_{st}(S\cup\{(u,v)\})$, then either $(u,v)$ does not belong to any shortest path from $s$ to $t$ and then $b_{stv}(S\cup\{(u,v)\}) = b_{stv}(S)$, or $(u,v)$ belongs to a newly added shortest path from $s$ to $t$ with the same weight and $b_{stv}(S\cup\{(u,v)\}) = \frac{\sigma_{stv}(S)+\delta}{\sigma_{st}(S)+\delta}>  \frac{\sigma_{stv}(S)}{\sigma_{st}(S)} = b_{stv}(S)$, where $\delta\geq 1$ is the number of shortest paths from $s$ to $t$ that pass through arc $(u,v)$ in $G(S\cup\{(u,v)\})$. In any case the statement holds.

We now show that $b_{stv}$ is submodular, that is for each pair of solutions to \MBI $S,T$ such that $S\subseteq T$ and for each node $u$ such that $(u,v)\not\in T\cup E$,
\begin{align*}
 b_{stv}(S\cup\{(u,v)\}) - b_{stv}(S) \geq b_{stv}(T\cup \{(u,v)\}) - b_{stv}(T).
\end{align*}
We analyze the following cases:
\begin{itemize}
  \item $d_{st}(S)>d_{st}(T)$. In this case, $b_{stv}(T\cup \{(u,v)\}) - b_{stv}(T) = 0$ since in any case $b_{stv}(T\cup \{(u,v)\}) = b_{stv}(T) = 1$. As $b_{stv}$ is monotone increasing, then $b_{stv}(S\cup\{(u,v)\}) - b_{stv}(S) \geq 0$.
  \item $d_{st}(S)=d_{st}(T)$.
  \begin{itemize}
   \item $d_{st}(S)>d_{st}(S\cup\{(u,v)\})$. In this case there exists a shortest path from $s$ to $t$ passing through edge $(u,v)$ in $G(S\cup\{(u,v))$ and the length of such path is strictly smaller that the distance from $s$ to $t$ in $G(S)$. Since $d_{st}(S)=d_{st}(T)$, such a path is a shortest path also in $G(T\cup\{(u,v)\})$ and its length is strictly smaller than $d_{st}(T)$. It follows that  $d_{st}(T)>d_{st}(T\cup\{(u,v)\})$ and $b_{stv}(T\cup\{(u,v)\}) = b_{stv}(S\cup\{(u,v)\}) = 1$. Moreover $b_{stv}(T)\geq b_{stv}(S)$. Therefore $b_{stv}(S\cup\{(u,v)\}) - b_{stv}(S) \geq b_{stv}(T\cup \{(u,v)\}) - b_{stv}(T)$.
   \item $d_{st}(S)=d_{st}(S\cup\{(u,v)\})$. In this case $d_{st}(T)=d_{st}(T\cup\{(u,v)\})$. Let us denote $b_{stv}(S)=\frac{\alpha}{\beta}$, then we have that $b_{stv}(T) = \frac{\alpha+\gamma}{\beta+\gamma}$, $b_{stv}(S\cup\{(u,v)\}) = \frac{\alpha+\delta}{\beta+\delta}$, and $b_{stv}(T\cup\{(u,v)\}) = \frac{\alpha+\gamma+\delta}{\beta+\gamma+\delta}$, where $\gamma$ and $\delta$ are the number of shortest paths between $s$ and $t$ in $G(T)$ that pass through arcs in $T\setminus S$ and arc $(u,v)$, respectively. The statement follows since $\frac{\alpha+\delta}{\beta+\delta}-\frac{\alpha}{\beta}\geq \frac{\alpha+\gamma+\delta}{\beta+\gamma+\delta}-\frac{\alpha+\gamma}{\beta+\gamma}$ for any $\alpha\leq \beta$, i.e. $\sigma_{stv}(S)\leq \sigma_{st}(S)$.\qed
  \end{itemize}
\end{itemize}

\end{proof}

\begin{corollary}
 Algorithm~\ref{alg:greedy} provides a $\left(1-\frac{1}{e}\right)$-approximation for the \MBI problem.
\end{corollary}

It is easy to compute the computational complexity of Algorithm \Greedy. Line~\ref{greedy1} iterates over all the numbers from $1$ to $k$. Then, in Line~\ref{greedy2}, all the nodes $u$ that are not yet neighbors of $v$ are scanned. The number of these nodes is clearly $O(n)$. Finally, in Line~\ref{greedy3}, for each node $u$ in Line~\ref{greedy2}, we add the edge $\{ u, v \}$ to the graph and compute the betweenness in the new graph. Since computing betweenness requires $O(n m)$ operations in unweighted graphs, the total running time of \Greedy is $O(k n^2 m)$. In Section~\ref{sec:dynamic-single} we show how to decrease this running time to $O(k n^3)$ by using a dynamic algorithm for the computation of betweenness centrality at Line~\ref{greedy3}.


\section{Dynamic algorithm for betweenness centrality of a single node}
\label{sec:dynamic-single}
\begin{wrapfigure}{r}{0.3\textwidth}
  \begin{center}
    \includegraphics[width=0.28\textwidth]{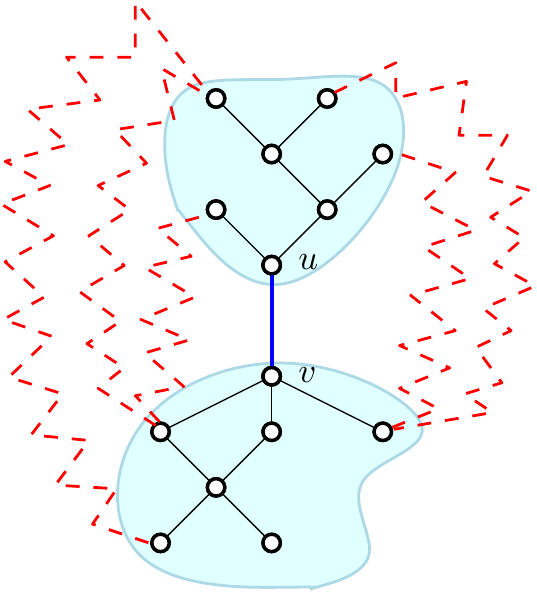}
  \end{center}
  \caption{Insertion of edge $(u,v)$ affects the betweenness of nodes lying in the old shortest paths (red).}
   \vspace{-3ex}
  \label{fig:old-paths}
\end{wrapfigure}
Algorithm~\ref{alg:greedy} requires to add edges to the graph and to recompute the betweenness centrality $b_v$ of node $v$ after each edge insertion. Instead of recomputing it from scratch every time, we use a dynamic algorithm. The idea is to keep track of information regarding the graph and just update the parts that have changed as a consequence of the edge insertion. 
As described in Section~\ref{sec:related}, several algorithms for updating betweenness centrality after an edge insertion have been proposed. However, these algorithms update the betweenness of \textit{all nodes}, whereas in Algorithm~\ref{alg:greedy} we are interested in the betweenness of a \textit{single node}. In this case, using an algorithm that recomputes the betweenness of all nodes would require a significant amount of superfluous operations. Let us consider the example shown in Figure~\ref{fig:old-paths}. 

The insertion of an edge $(u,v)$ does not only affect the betweenness of the nodes lying in the new shortest paths, but also that of the nodes lying in the old shortest paths between affected sources and affected targets (represented in red).
Indeed, the fraction of shortest paths going through these nodes (and therefore their betweenness) has decreased as a consequence of the new insertion. Therefore, algorithms for updating the betweenness of all nodes have to walk over each old shortest path between node pairs whose distance has changed. 
However, we will show that if we are only interested in the betweenness of one particular node $x$, we can simply update the distances (and number of shortest paths) and check which of these updates affect the betweenness of $x$. Section~\ref{sec:dynamic} describes our new dynamic algorithm for updating the betweenness of a single node after an edge insertion. Notice that the algorithm could be used in any context where one needs to keep track of the betweenness of a single node after an edge insertion (or weight decrease) and not only for the betweenness improvement. Since our new algorithm builds on a recent dynamic betweenness algorithm called \textsf{iBet}~\cite{DBLP:conf/sea/BergaminiMOS17}, we first describe \textsf{iBet} in Section~\ref{sec:quinca} and then explain how this can be modified to recompute the betweenness of a single node in Section~\ref{sec:dynamic}.
\subsection{\textsf{iBet} algorithm for updating the betweenness of all nodes}
\label{sec:quinca}
\textsf{iBet}~\cite{DBLP:conf/sea/BergaminiMOS17} updates the betweenness of all nodes after an edge insertion or an edge weight decrease. 
Just as Brandes's algorithm~\cite{Brandes01betweennessCentrality}, \textsf{iBet} is composed of two steps: a step where the pairwise distances and number of shortest paths are computed, and a step where the actual betweenness values are found. 

Let us assume a new edge $(u, v)$ with weight $\omega'_{u, v}$ is inserted into the graph, or that the weight of an existing edge $(u, v) \in E$ is decreased and set to a new value $\omega'_{u, v}$. 
Then, let us name \textit{affected pairs} the node pairs $(s, t)$ such that $d_{st} \geq  d_{su} + \omega'_{u, v} + d_{vt}$. Notice that these are the nodes for which either $(u, v)$ creates a shortcut (decreasing the distance), or creates one or more new shortest paths of the same length as the old distance.
 Also, let the \textit{affected sources} of a node $t$ be the set $S(t)$ of nodes $\{s \in V : d_{st} \geq d_{su} + \omega'_{u, v} + d_{vt}\}$ and let the set $T(s)$ of \textit{affected targets} of $s$ be $T(s) = \{t \in v : d_{st} \geq  d_{su} + \omega'_{u, v} + d_{vt}\}$.


In~\cite{DBLP:conf/sea/BergaminiMOS17} it was proven that if $(s,t)$ is an affected node pair, then $s \in S(v)$ and $t \in T(u)$. This allows us to reduce the search space of the affected pairs to the nodes whose distance to $v$ or from $u$ has changed (or their number of shortest paths). Thus, a first idea would be to identify the set $S(v)$ and the set $T(u)$, which can be done with two pruned BFSs, rooted in $u$ and $v$, respectively. For each node $s \in S(v)$ and each node $t \in T(u)$, we can compare the old distance $d_{st}$ with the one of a path going through edge $(u, v)$, namely $d_{su} + \omega'_{u, v} + d_{vt}$, and update the distance and number of shortest paths accordingly. However, \textsf{iBet} is more efficient than this. Let a \textit{predecessor} in a shortest path from $v$ to $t$ be any node $x$ such that $(x, t) \in E$ and $d_{vt} = d_{vx}+d_{xt}$, and let us denote this as $x \in P_v(t)$. Then, the following lemma holds.
\begin{lemma}{\cite{DBLP:conf/sea/BergaminiMOS17}}
\label{lemma:ibet}
Let $\ t \in V$ be any node and $x \in P_v(t)$ be a predecessor of $t$ in the shortest paths from $v$. Then, $S(t) \subseteq S(x)$.
\end{lemma}
\begin{wrapfigure}{r}{0.2\textwidth}
  \begin{center}
    \includegraphics[width=0.15\textwidth]{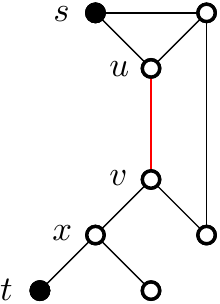}
  \end{center}
  \vspace{-1ex}
  \caption{Insertion of edge $(u,v)$ creates a shortcut between $s$ and $t$, but also between $s$ and $x$.}
  \label{fig:pred}
    \vspace{-3ex}
\end{wrapfigure}
Figure~\ref{fig:pred} explains this concept. The insertion of edge $(u,v)$ creates a shortcut between $s$ and $t$, making $(s,t)$ an affected pair. Similarly, the new edge creates a shortcut between $s$ and each predecessor of $t$ in the shortest path from $v$, i.e. $x$. Lemma~\ref{lemma:ibet} basically tells us that we do not need to check all pairs $(s, t)$ such that $s \in S(v)$ and $t \in T(u)$. On the contrary, for a given target $t$, we only need to check the nodes $s \in S(x)$, where $x$ is any node in $P_v(t)$. If we process the targets in increasing order of distance from $v$, this set will become smaller and smaller as we go down the BFS tree, saving unnecessary comparisons.

Clearly, what we described is only the update of the augmented APSP. After this, \textsf{iBet} also needs to update the betweenness scores of all the nodes that lie in some old or new shortest path between an affected pair. Since this part is not necessary when updating the betweenness of a single node, we will not describe this and refer the reader to~\cite{DBLP:conf/sea/BergaminiMOS17} for more details.

Since the augmented APSP update of \textsf{iBet} was shown to be significantly faster than all existing algorithms, we use it as a building block for our incremental algorithm for the betweenness centrality of a single node, described in Section~\ref{sec:dynamic}.

\subsection{New dynamic algorithm for the betweenness of a single node}
\label{sec:dynamic}
\textsf{iBet} stores the pairwise distances $d_{st}$ and the number of shortest paths $\sigma_{st}$ for each $s, t \in V$. When computing the betweenness of a specific node $x$, we also need the number $\sigma_{stx}$ of shortest paths between $s$ and $t$ that go through $x$. Then, we can compute betweenness by using its definition given in Eq. (\ref{def:bc}). 
In the following, we will assume that the graph $G$ is unweighted and connected, but the algorithm can be easily extended to weighted and disconnected graphs, in a way analogous to what has been done in~\cite{DBLP:conf/sea/BergaminiMOS17}. Our algorithm can be divided in two phases: an \textit{initialization} phase, where pairwise distances, $\sigma_{(\cdot, \cdot)}$ and $\sigma_{(\cdot, \cdot) x}$ are computed and stored, and an \textit{update} phase, where the data structures and the betweenness of node $x$ are updated as a consequence of the edge insertion.
\subsubsection{Initialization} 
The initialization can be easily done by running a Single-Source Shortest Path (SSSP) from each node, as in the first phase of Brandes's algorithm for betweenness centrality~\cite{Brandes01betweennessCentrality}. While computing distances from a source node $s$ to any other node $t$, we set the number $\sigma_{st}$ of shortest paths between $s$ and $t$ to the sum $\sum \sigma_{sp}$ over all predecessors $p$ in the shortest paths from $s$ (and we set $\sigma_{ss} = 1$). This can be done for a node $s$ in $O(m)$ in unweighted graphs and in $O(m + n \log n)$ in weighted graphs (the cost of running a BFS or Dijkstra, respectively). Instead of discarding this information after each SSSP as in Brandes's algorithm, we store both the distances $d_{(\cdot, \cdot)}$ and the numbers of shortest paths $\sigma_{(\cdot, \cdot)}$ in a matrix. After this, we can compute the number $\sigma_{(\cdot, \cdot) x}$ of shortest paths going through $x$. For each node pair $(s,t)$, $\sigma_{stx}$ is equal to $\sigma_{sx} \cdot \sigma_{xt}$ if $d_{st} = d_{sx} + d_{xt}$, and to 0 otherwise. The betweenness $b_x$ of $x$ can then be computed using the definition given in Eq. (\ref{def:bc}).
This second part can be done in $O(n^2)$ time by looping over all node pairs. Therefore the total running time of the initialization is $O(nm)$ for unweighted graphs and $O(n(m + n\log n))$ for weighted graphs, and the memory requirement is $O(n^2)$, since we need to store three matrices of size $n \times n$ each.
\subsubsection{Update}
The update works in a way similar to \textsf{iBet} (see Section~\ref{sec:quinca}), with a few differences. Algorithm~\ref{algo:overview} gives an overview of the algorithm for betweenness update for a single node $x$, whereas Algorithm~\ref{algo:updateSigmaGR} and Algorithm~\ref{algo:updateSigmaEQ}
describe the update $\sigma_{(\cdot, \cdot)}$ and $\sigma_{(\cdot, \cdot) x}$ when $d_{st} > d_{su} + \omega'_{uv} + d_{vt}$ and when $d_{st} = d_{su}+\omega'_{uv}+d_{vt}$, respectively.

 Algorithm~\ref{algo:overview} shares its structure with \textsf{iBet}. In Lines~\ref{initialization1}-\ref{initialization2}, after setting the new distance between $u$ and $v$, also $\sigma_{uv}$ and $\sigma_{uvx}$ are updated with either \textsf{updateSigmaGR} or \textsf{updateSigmaEQ}. Then, just like in \textsf{iBet}, the affected sources are identified with a pruned BFS rooted in $u$ on $G$ transposed (function \textsf{findAffectedSources}).

Then, a (pruned) BFS rooted in $v$ is started to find the affected targets for $u$ (Lines~\ref{while-start}-\ref{while-end}). In Lines~\ref{visit-neigh}-\ref{while-end}, the neighbors $w$ of the affected target $t$ are visited and, if they are also affected (i.e. $d_{uw} \geq \omega'_{uv} + d_{vw}$), they are enqueued. Also, $t$ is stored as the predecessor of $w$ (Line~\ref{while-end}).
In Lines~\ref{ifGEQ}-\ref{endIfGEQ}, for each affected node pair $(s, t)$, we first subtract the old contribution $\sigma_{stx}/ \sigma_{st}$ from the betweenness of $x$, then we recompute $d_{st}$,  $\sigma_{st}$ and $\sigma_{stx}$ with either \textsf{updateSigmaGR} or \textsf{updateSigmaEQ}, and finally we add the new contribution $\sigma'_{stx}/ \sigma'_{st}$ to $b_x$. Notice that, if $x$ did not lie in any shortest path between $s$ and $t$ before the edge insertion, $\sigma_{stx} = 0$ and therefore $b_x$ is not decreased in Line~\ref{decreaseBC}. Analogously, if $x$ is not part of a shortest path between $s$ and $t$ after the insertion, $b_x$ is not increased in Line~\ref{increaseBC}.

In the following, we analyze \textsf{updateSigmaGR} and \textsf{updateSigmaEQ} separately.
\begin{algorithm2e}[tbh]
\begin{footnotesize}
\SetKwInput{Proc}{Algorithm}
\Proc{Incremental betweenness}
\SetKwInOut{Input}{Input}\SetKwInOut{Output}{Output}
\SetKwInOut{Assume}{Assume}
\SetKwFunction{insert}{insert}
\SetKwFunction{findAffectedSources}{findAffectedSources}
\SetKwFunction{updateGR}{updateSigmaGR}
\SetKwFunction{updateEQ}{updateSigmaEQ}
\SetKwData{visited}{visited}
\SetKwData{Pred}{$P_v$}
\SetKwData{source}{$S$}
\Input{Graph $G=(V, E)$, edge update $(u,v,\omega'_{uv})$, pairwise distances $d_{(\cdot, \cdot)}$, numbers $\sigma_{(\cdot, \cdot)}$ of shortest paths, numbers $\sigma_{(\cdot, \cdot) x}$ of shortest paths through $x$, betweenness value $b_x$ of $x$}
\Output{Updated  $d_{(\cdot, \cdot)}$, $\sigma_{(\cdot, \cdot)}$, $\sigma_{(\cdot, \cdot) x}$ and $b_x$}
\Assume{boolean $vis(v)\text{ is false},\ \forall v \in V$}
\If{$d_{uv} \geq \omega'_{uv}$}{   
	\If{$d_{uv} > \omega'_{uv}$}{ \label{initialization1} 
		$d_{uv} \leftarrow \omega'_{uv}$\;
		$\sigma_{uv}, \sigma_{uvx}\leftarrow$\updateGR{$G, (u,v), d, \sigma, \sigma_x$}\;
	}\Else{
		$\sigma_{uv}, \sigma_{uvx}\leftarrow$\updateEQ{$G, (u,v), d, \sigma, \sigma_x$}\;
	} \label{initialization2} 
  $S(v) \leftarrow$ \findAffectedSources{$G, (u,v), d$}\; \label{aff-sourcesGEQ}
  $Q \leftarrow \emptyset$\;
  $P(v) \leftarrow v$\;
  $Q.push(v)$\;
  $vis(v) \leftarrow \text{true}$\;
  \While {$Q.length() > 0$}{ 
  \label{while-start}
    $t = Q.front()$\; 
    \ForEach {$s \in \source(P(t))$} { 
        \label{neighbors-begin}
      \If {$d_{st} \geq d_{su} + \omega'_{uv} + d_{vt}$} { \label{ifGEQ}
      		\If{$x \neq s$ and $x \neq t$} {
      	     $b_x \leftarrow b_x - \sigma_{stx}/ \sigma_{st}$\; \label{decreaseBC}
	     }
              \If {$d_{st} > d_{su} + \omega'_{uv} + d_{vt}$} { 
              		$\sigma_{st}, \sigma_{stx}\leftarrow$\updateGR{$G, (u,v), d, \sigma, \sigma_x$}\; \label{updateGR}
			$d_{st} \leftarrow d_{su} + \omega'_{uv} + d_{vt}$\;
     	 	}
		\Else {
			$\sigma_{st}, \sigma_{stx}\leftarrow$\updateEQ{$G, (u,v), d, \sigma, \sigma_x$}\; \label{updateEQ}
		} 
		\If{$x \neq s$ and $x \neq t$} {
			$b_x \leftarrow b_x + \sigma_{stx}/ \sigma_{st}$\; \label{increaseBC}
		}
       		 \If{$t \neq v$} {
        			$S(t).insert(s)$\; \label{endIfGEQ}
        		}
      }  
    } 
    \ForEach {$w$ s.t. $(t, w) \in E$} { 
    \label{visit-neigh}
      \If {not $vis(w)$ and $d_{uw} \geq \omega'_{uv} + d_{vw}$}{ 
      \label{distanceGEQ}
        $Q.push(w)$\; 
          $vis(w) \leftarrow \text{true}$\;
          $P(w) \leftarrow t$\;   \label{while-end}
      } 
    } 
  } 
 }
\end{footnotesize}
\caption{Update of $b_x$ after an edge insertion}
\label{algo:overview}
\end{algorithm2e}

\begin{figure}[tbh]
\begin{center}
\includegraphics[width = 0.25\textwidth]{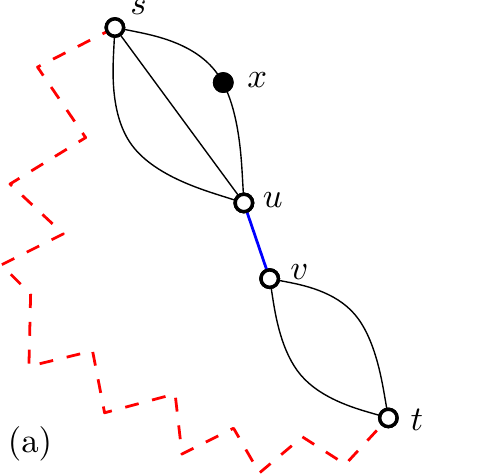}\hspace{3ex}
\includegraphics[width = 0.25\textwidth]{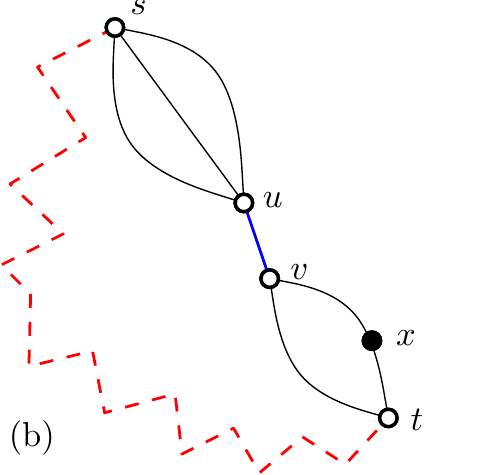}\hspace{3ex}
\includegraphics[width = 0.25\textwidth]{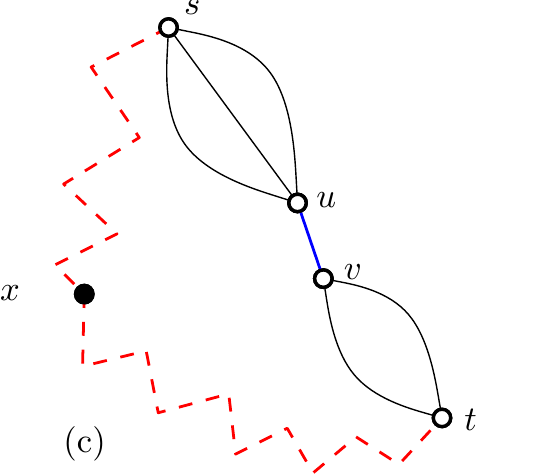}
\caption{Possible positions of $x$ with respect to the new shortest paths after the insertion of edge $(u, v)$. On the left, $x$ lies between the source $s$ and $u$. In the center, $x$ lies between $v$ and the target $t$. On the right, $x$ does not lie on any new shortest path between $s$ and $t$.} 
\label{fig:casesGR}
\end{center}
\end{figure}

\paragraph*{\textsf{UpdateSigmaGR}}
Let us consider the case $d_{st} > d_{su} + \omega'_{uv} + d_{vt}$. In this case, all the old shortest paths are discarded, as they are not shortest paths any longer, and all the new shortest paths go through edge $(u,v)$. Therefore, we can set the new number $\sigma'_{st}$ of shortest paths between $s$ and $t$ to $\sigma_{su}\cdot \sigma_{vt}$. Since all old shortest paths should be discarded, also $\sigma_{stx}$ depends only on the new shortest paths and not on whether $x$ used to lie in some shortest paths between $s$ and $t$ before the edge insertion.
Depending on the position of $x$ with respect to the new shortest paths, we can define three cases, depicted in Figure~\ref{fig:casesGR}. In Case \textsf{(a)} (left), $x$ lies in one of the shortest paths between $s$ and $u$. This means that it also lies in some shortest paths between $s$ and $t$. In particular, the number of these paths $\sigma'_{stx}$ is equal to $\sigma_{sux} \cdot \sigma_{vt}$. Notice that no shortest paths between $s$ and $u$ can be affected (see~\cite{DBLP:conf/sea/BergaminiMOS17}) and therefore $\sigma_{sux} = \sigma'_x(s,u)$. In Case \textsf{(b)} (center), $x$ lies in one of the shortest paths between $v$ and $t$. Analogously to Case 1, the new number of shortest paths between $s$ and $t$ going through $x$ is $\sigma'_{stx} = \sigma_{su} \cdot \sigma_{vtx}$. Notice that Case \textsf{(a)} and Case \textsf{(b)} cannot both be true at the same time. In fact, if $d_{su} = d_{sx} + d_{xu}$ and $d_{vt} = d_{vx} + d_{xt}$, we would have that $d'_{st} = d_{su} + \omega'_{uv} + d_{vt} = d_{sx} + d_{xu} + \omega'_{uv} + d_{vx} + d_{xt} > d_{sx} + d_{xt}$, which is impossible, since $d'_{st}$ is the shortest-path distance between $s$ and $t$. Therefore, at least one among $\sigma_{sux}$ and $\sigma_{vtx}$ must be equal to 0. Finally, in Case \textsf{(c)} (right), $\sigma_{sux}$ and $\sigma_{vtx}$ are both equal to 0, meaning that $x$ does not lie on any new shortest path between $s$ and $t$. Once again, this is independent on whether $x$ lied in an old shortest path between $s$ and $t$ or not. Algorithm~\ref{algo:updateSigmaGR} shows the computation of $\sigma'_{st}$ and $\sigma'_{stx}$. Notice that, in the computation of $\sigma'_{stx}$, the first addend is greater than zero only in Case \textsf{(a)} and the second only in Case \textsf{(b)}.

\begin{algorithm2e}[tb]
\begin{footnotesize}
\SetKwInput{Proc}{Algorithm}
\Proc{\textsf{UpdateSigmaGR}}
\SetKwInOut{Input}{Input}\SetKwInOut{Output}{Output}
\SetKwInOut{Assume}{Assume}
\SetKwFunction{insert}{insert}
\SetKwData{visited}{visited}
\SetKwData{Pred}{$P_v$}
\SetKwData{source}{$S$}
\Input{Graph $G=(V, E)$, edge insertion $(u,v)$, pairwise distances $d_{(\cdot, \cdot)}$, numbers $\sigma_{(\cdot, \cdot)}$ of shortest paths, numbers $\sigma_{(\cdot, \cdot) x}$ of shortest paths through $x$}
\Output{Updated  $\sigma'_{st}$, $\sigma'_{stx}$}
$\sigma'_{st} \leftarrow \sigma_{su} \cdot \sigma_{vt}$\;
$\sigma'_{stx} \leftarrow \sigma_{sux} \cdot \sigma_{vt} + \sigma_{su} \cdot \sigma_{vtx}$\;
\Return $\sigma'_{st}$, $\sigma'_{stx}$\;
\end{footnotesize}
\caption{Update of $\sigma_{st}$ and $\sigma_{stx}$ when $(u,v)$ creates new shortest paths of length smaller than $d_{st}$}
\label{algo:updateSigmaGR}
\end{algorithm2e}

\paragraph*{\textsf{UpdateSigmaEQ}}
Let us now consider the case $d_{st} = d_{su} + \omega'_{uv} + d_{vt}$. Here all the old shortest paths between $s$ and $t$ are still valid and, in addition to them, new shortest paths going through $(u,v)$ have been created. Therefore, the new number of shortest paths $\sigma'_{st}$ is simply $\sigma_{st} + \sigma_{su} \cdot \sigma_{vt}$. Notice that we never count the same path multiple times, since all new paths go through $(u,v)$ and none of the old paths does. Also all old shortest paths between $s$ and $t$ through $x$ are still valid, therefore $\sigma'_{stx}$ is given by the old $\sigma_{stx}$ plus the number of new shortest paths going through both $x$ and $(u, v)$. This number can be computed as described for \textsf{updateSigmaGR} according to the cases of Figure~\ref{fig:casesGR}. Algorithm~\ref{algo:updateSigmaEQ} shows the computation of $\sigma'_{st}$ and $\sigma'_{stx}$.

\begin{algorithm2e}[tb]
\begin{footnotesize}
\SetKwInput{Proc}{Algorithm}
\Proc{\textsf{UpdateSigmaEQ}}
\SetKwInOut{Input}{Input}\SetKwInOut{Output}{Output}
\SetKwInOut{Assume}{Assume}
\SetKwFunction{insert}{insert}
\SetKwData{visited}{visited}
\SetKwData{Pred}{$P_v$}
\SetKwData{source}{$S$}
\Input{Graph $G=(V, E)$, edge insertion $(u,v)$, pairwise distances $d_{(\cdot, \cdot)}$, numbers $\sigma_{(\cdot, \cdot)}$ of shortest paths, numbers $\sigma_{(\cdot, \cdot) x}$ of shortest paths through $x$}
\Output{Updated  $\sigma'_{st}$, $\sigma'_{stx}$}
$\sigma'_{st} \leftarrow \sigma_{st} + \sigma_{su} \cdot \sigma_{vt}$\;
$\sigma'_{stx} \leftarrow \sigma_{stx} + \sigma_{sux} \cdot \sigma_{vt} + \sigma_{su} \cdot \sigma_{vtx}$\;
\Return $\sigma'_{st}$, $\sigma'_{stx}$\;
\end{footnotesize}
\caption{Update of $\sigma_{st}$ and $\sigma_{stx}$ when $(u,v)$ creates new shortest paths of length equal to $d_{st}$}
\label{algo:updateSigmaEQ}
\end{algorithm2e}
\subsection{Time complexities}
\subsubsection{Dynamic betweenness algorithm}
\label{sec:complexity}
Let us define the \textit{extended size} $||A||$ of a set of nodes $A$ as the sum of the number of nodes in $A$ and the number of edges that have a node of $A$ as their endpoint. Then, the following proposition holds.
\begin{proposition}
The running time of Algorithm~\ref{algo:overview} for updating the betweenness of a single node after an edge insertion $(u, v)$ is $\Theta(||S(v)|| + ||T(u)|| + \sum_{y \in T(u)} |S(P(y))|)$.
\end{proposition}
\begin{proof}
The function \texttt{findAffectedSources} in Line~\ref{aff-sourcesGEQ} identifies the set of affected sources starting a BFS in $v$ and visiting only the nodes $s$ such that $d_{su} + \omega'_{uv} + d_{vt} \leq d_{st}$. This takes $\Theta(||S(v)||)$, since this partial BFS visits all nodes in $S(v)$ and their incident edges.
Then, the while loop of Lines~\ref{while-start} -~\ref{while-end} (excluding the part in Lines~\ref{neighbors-begin} -~\ref{endIfGEQ}) identifies all the affected targets $T(u)$ with a partial BFS. This part requires $\Theta(||T(u)||)$ operations, since all affected targets and their incident edges are visited. 
In Lines~\ref{neighbors-begin} -~\ref{endIfGEQ}, for each affected node $t \in T(u)$, all the affected sources of the predecessor $P(t)$ of $t$ are scanned. This part requires in total $\Theta(\sum_{t \in T(u)} |S(P(t))|)$ operations, since for each node in $S(P(t))$, Lines~\ref{ifGEQ} -~\ref{endIfGEQ} require constant time.
\end{proof}
Notice that, since $S(P(y))$ is $O(n)$ and both $||T(u)||$ and $||S(v)||$ are $O(n+m)$, the worst-case complexity of Algorithm~\ref{algo:overview} is $O(n^2)$ (assuming $m = \Omega(n)$). This matches the worst-case running time of the augmented APSP update of \textsf{iBet}. However, notice that \textsf{iBet} needs a second step to update the betweenness of all nodes, which is more expensive and requires $\Theta(nm)$ operations in the worst case.
Also, this introduces a contrast between the static and the incremental case: Whereas the static computation of one node's betweenness has the same complexity as computing it for all nodes (at least no algorithm for computing it for one node faster than computing it for all nodes exists so far), in the incremental case the betweenness update of a single node can be done in $O(n^2)$, whereas there is no algorithm faster than $O(nm)$ for the update of all nodes.

\subsubsection{Greedy algorithm for betweenness maximization}
We can improve the running time of Algorithm \Greedy by using the dynamic algorithm for betweenness centrality instead of the recomputation from scratch. In fact, at Line~\ref{greedy3} of Algorithm \Greedy, we add an edge $\{ u, v \}$ to the graph and compute the betweenness in the new graph, for each node $u$ in $V\setminus N_v(S)$. If we compute the betweenness by using the from-scratch algorithm, this step requires $O(n m)$ and this leads to an overall complexity of $O(k n^2 m)$. At Line~\ref{greedy3} , instead of recomputing betweenness on the new graph from scratch, we can use Algorithm~\ref{algo:overview}. As we proved previously, its worst-case complexity is $O(n^2)$. This leads to an overall worst-case complexity of $O(k n^3)$ for \Greedy. However, in Section~\ref{sec:experiments} we will show that \Greedy is actually much faster in practice.

\section{Experimental evaluation}
In our experiments, we evaluate the performance of \Greedy both in terms of quality of the solution found (Section~\ref{sec:experiments_greedy}) and in terms of its running time (Section~\ref{sec:experiments}). All algorithms compared in our experiments are implemented in C++, building on the open-source NetworKit~\cite{Staudt2014} framework. 
The experiments were done on a machine equipped with 256 GB RAM and a 2.7 GHz Intel Xeon CPU E5-2680 having 2 sockets with 8 cores each. To make the comparison with previous work more meaningful, we use only one of the 16 cores.
The machine runs 64 bit SUSE Linux and we compiled our code with g++-4.8.1 and OpenMP~3.1. 

For our experiments, we consider a set of real-world networks belonging to different domains, taken from SNAP~\cite{snapnets}, KONECT~\cite{DBLP:conf/www/Kunegis13}, Pajek \cite{pajek}, and the 10th DIMACS Implementation Challenge~\cite{BaderMSSKW14benchmarking}. 
The properties of the networks are reported in Table~\ref{table:directed} (directed graphs) and  in Table~\ref{table:undirected} (undirected graphs).
\subsection{Solution quality}
\label{sec:experiments_greedy}
In this section we evaluate \Greedy in terms of accuracy and we compare it both with the optimum and with some alternative baselines.

To speed up the computation of \Greedy (and therefore to target larger graphs), we do not recompute betweenness from scratch in Line~\ref{greedy3} of Algorithm~\ref{alg:greedy}, but we use the dynamic algorithm described in Section~\ref{sec:dynamic-single}. Notice that this does not affect the solution found by the algorithm, only its running time, which is reported in Section~\ref{sec:experiments}. Since computing the optimum by examining all possible $k$-tuples would be too expensive even on small graphs, we use an Integer Programming (IP) formulation, described in the following paragraph.


\subsubsection{IP formulation for \MBI on directed graphs}
Let $S$ be a solution to an instance of \MBI. Given a node $v$, we define a variable $x_{u}$ for each node $u \in V\setminus (N_v \cup \{v\})$
\[
 x_{u} = \begin{cases}
        1&\mbox{if } (u,v)\in S\\
        0 &\mbox{otherwise.}
       \end{cases} 
\]
We define a variable $y_{st}$ for each $s,t\in V\setminus\{v\}$, $s\neq t$.
\[
  y_{st} =  \begin{cases}
        1&\mbox{If all shortest paths from } s \mbox{ to } t \mbox{ in } G(S) \mbox{ pass through node } v\\
        0 &\mbox{otherwise.}
       \end{cases} 
\]

For each pair of nodes $s,t\in V\setminus\{v\}$, $s\neq t$, we denote by $A(s,t)$ the set of nodes $u$ not in $N_v$ such that all the shortest paths between $s$ and $t$ in $G(\{(u,v)\})$ pass through edge $(u,v)$ and hence through node $v$. Note that in this case, $d_{st}>d_{st}(\{(u,v)\})$ and hence $A(s,t)$  is defined as $A(s,t) = \{u~|~d_{st}>d_{st}(\{(u,v)\})\}$. Set $B(s,t)$ is defined as the set of nodes $u$ not in $N_v$ such that at least a shortest path between $s$ and $t$ in $G(\{(u,v)\})$ does not pass through edge $(u,v)$ and hence $B(s,t) = V \setminus (A(s,t) \cup N_v)$.
We denote by $\bar{\sigma}_{stv}(u)$ the number of shortest paths from $s$ to $t$ in $G(\{(u,v)\})$ passing through edge $(u,v)$.

The following non linear formulation solves the \MBI problem:
  \begin{align}
  \max &\sum_{\substack{s,t\in V\\s\neq t;s,t\neq v}} \left((1-y_{st})\frac{\sigma_{stv} + \sum_{u \in B(s,t)}x_u\bar{\sigma}_{stv}(u) }{\sigma_{st} + \sum_{u \in B(s,t)}x_u\bar{\sigma}_{stv}(u)} + y_{st} \right) \label{NLP:obj}\\
  \mbox{subject to}& \sum_{u\in A(s,t)}x_{u}\geq y_{st},& s,t\in V\setminus\{v\}, s\neq t\label{NLP:bet:one}\\
  &\sum_{u\in V\setminus (N_v \cup \{v\})} x_{u} \leq k,\nonumber\\
             & x_{u},y_{st}\in\{0,1\}&s\in V\setminus\{v\}, t\in V\setminus\{v,s\}\nonumber
 \end{align}

Let us consider a solution $S$ to the above formulation. In the case that $y_{st} = 1$, for some pair of nodes $s,t\in V\setminus\{v\}$, $s\neq t$, then Constraint~\eqref{NLP:bet:one} implies that, for at least a node $u\in A(s,t)$, variable $x_u$ must be set to 1 and hence all the shortest paths between $s$ and $t$ in $G(S)$ pass through $v$. In this case, the term corresponding to pair $(s,t)$ in the objective function~\eqref{NLP:obj} is correctly set to be equal to 1.

If $y_{st} = 0$ and $x_u = 0$, for each $u\in A(s,t)$, then the number of shortest paths between $s$ and $t$ in $G(S)$ passing through $v$ is equal to $\sigma_{stv} + \sum_{u \in B(s,t)}x_u\bar{\sigma}_{stv}(u) $ and the overall number of shortest paths between $s$ and $t$ in $G(S)$ is equal to $\sigma_{st} + \sum_{u \in B(s,t)}x_u\bar{\sigma}_{stv}(u)$. In this case, the term corresponding to pair $(s,t)$ in the objective function~\eqref{NLP:obj} is correctly set to be equal to $\frac{\sigma_{stv} + \sum_{u \in B(s,t)}x_u\bar{\sigma}_{stv}(u) }{\sigma_{st} + \sum_{u \in B(s,t)}x_u\bar{\sigma}_{stv}(u)}$.

Note that, $\frac{\sigma_{stv} + \sum_{u \in B(s,t)}x_u\bar{\sigma}_{stv}(u) }{\sigma_{st} + \sum_{u \in B(s,t)}x_u\bar{\sigma}_{stv}(u)}\leq 1$ and therefore a solution in which  $y_{st} = 0$ and $x_u = 1$, for some $u\in A(s,t)$ is at least as good as a solution in which $y_{st}$ is set to 1 instead of 0 and the other variables are unchanged. Hence, we can assume without loss of generality that the case in which $y_{st} = 0$ and $x_u = 1$, for some $u\in A(s,t)$, cannot occur in an optimal solution.



We solve the program with the Mixed-Integer Nonlinear Programming Solver Couenne~\cite{couenne} and measure the approximation ratio of the greedy algorithm on three types of randomly generated directed networks, namely directed Preferential Attachment (in short, \texttt{PA})~\cite{BBCR03}, Copying (in short, \texttt{COPY})~\cite{KRRSTU00}, Compressible Web (in short, \texttt{COMP})~\cite{CKLPR09}. For each graph type, we generate 5 different instances with the same size. We focus our attention on twenty vertices $v$, which have been chosen on the basis of their original betweenness ranking. In particular, we divide the list of vertices, sorted by their original ranking, in four intervals, and choose five random vertices uniformly at random in each interval. In each experiment, we add $k = \{ 1, 2, ..., 7\}$ edges. We evaluate the quality of the solution produced by the greedy algorithm by measuring its approximation ratio and we report the results in Table~\ref{table:opt}.
\begin{table}[h]
\caption{Comparison between the \Greedy algorithm and the optimum. The first three columns report the type and size of the graphs; the fourth column reports the approximation ratio.}{

  \begin{tabular}{ | l | r | r | c |}
    \hline
Graph	&	Nodes	&	Edges	&	Min. approx. ratio \\
\hline
\texttt{PA}	& 100	&	130	& 1\\ 
\texttt{COPY}	&	100	&	200	& 0.98  \\ 
\texttt{COMP}	&	100	&	200	& 0.98 \\ 
\texttt{COMP}	&	100	&	500	& 0.96 \\ 
\hline

  \end{tabular}
}
 \label{table:opt}
\end{table}
The experiments clearly show that the experimental approximation ratio is by far better than the theoretical one proven in the previous section. In fact, in all our tests, the experimental ratio is always greater than 0.96.

\subsubsection{Results for real-world directed networks}
\begin{figure}[htb]
\begin{center}
\includegraphics[width = 0.49\textwidth]{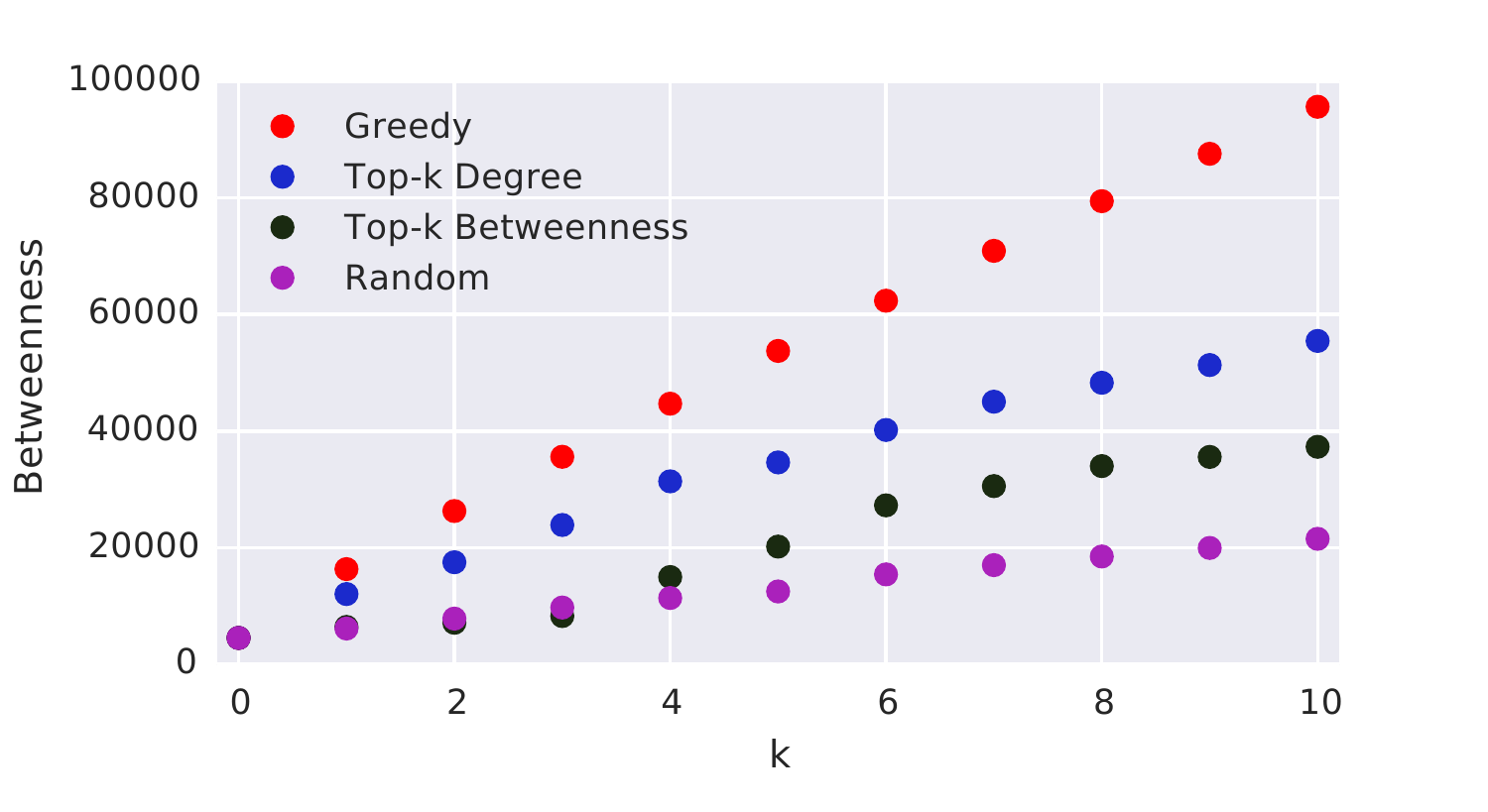}
\includegraphics[width = 0.49\textwidth]{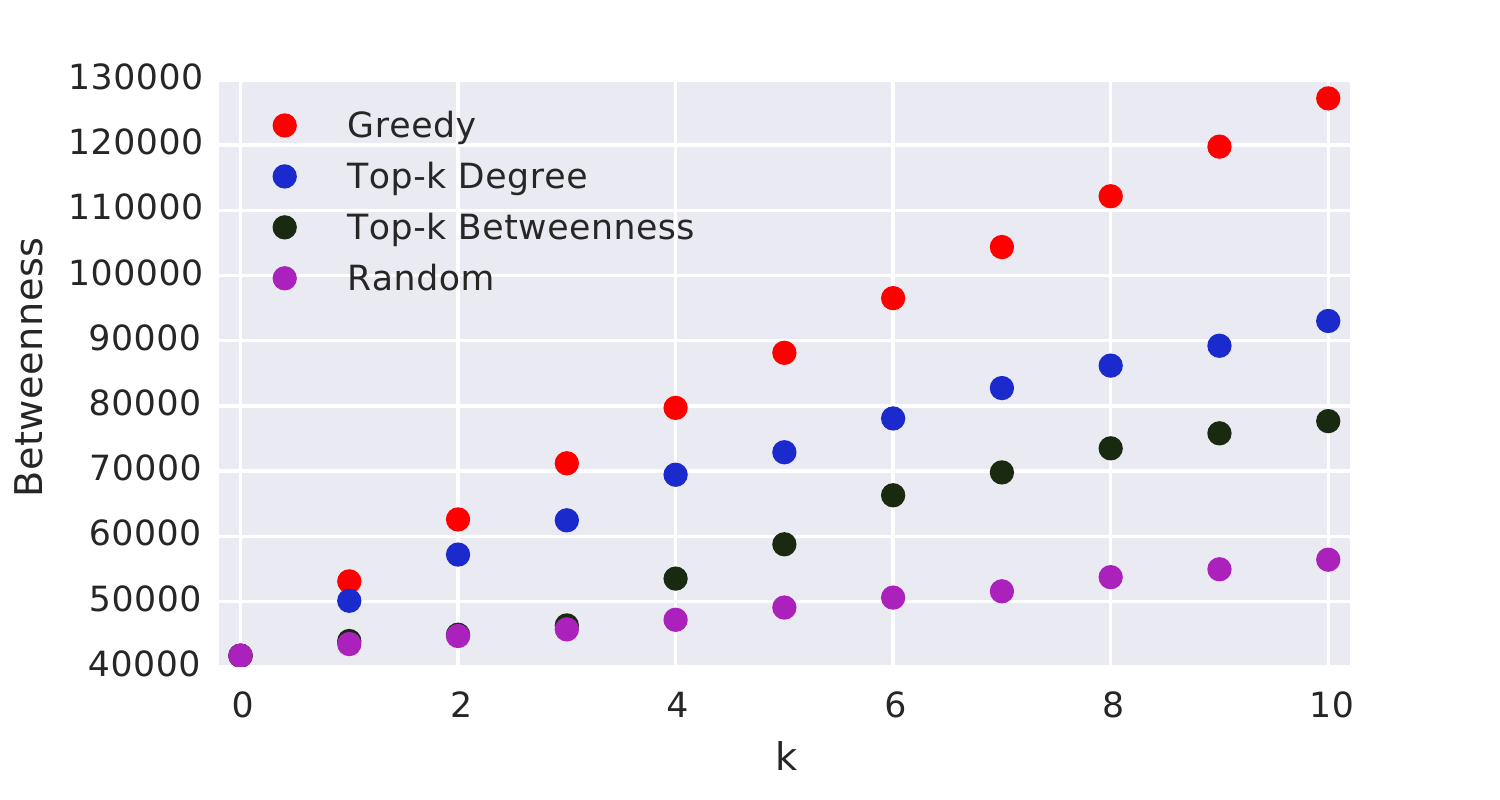}
\caption{Betweenness of the pivot as a function of the number $k$ of inserted edges for the four heuristics. The plots refer to two different pivots in the \texttt{munmun-digg-reply} graph.}
\label{fig:bc_munmun}
\end{center}
\end{figure}
\begin{figure}[htb]
\begin{center}
\includegraphics[width = 0.49\textwidth]{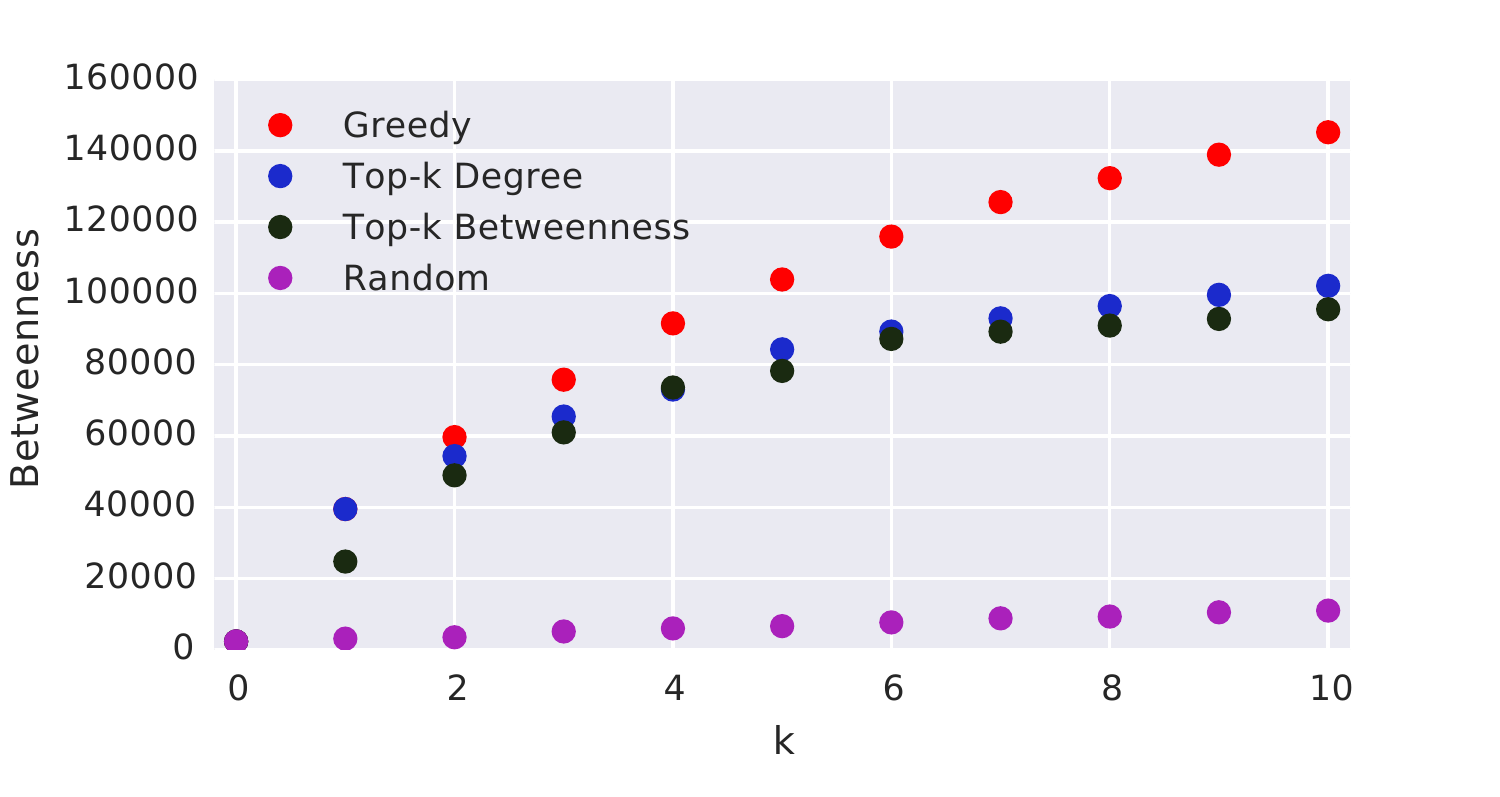}
\includegraphics[width = 0.49\textwidth]{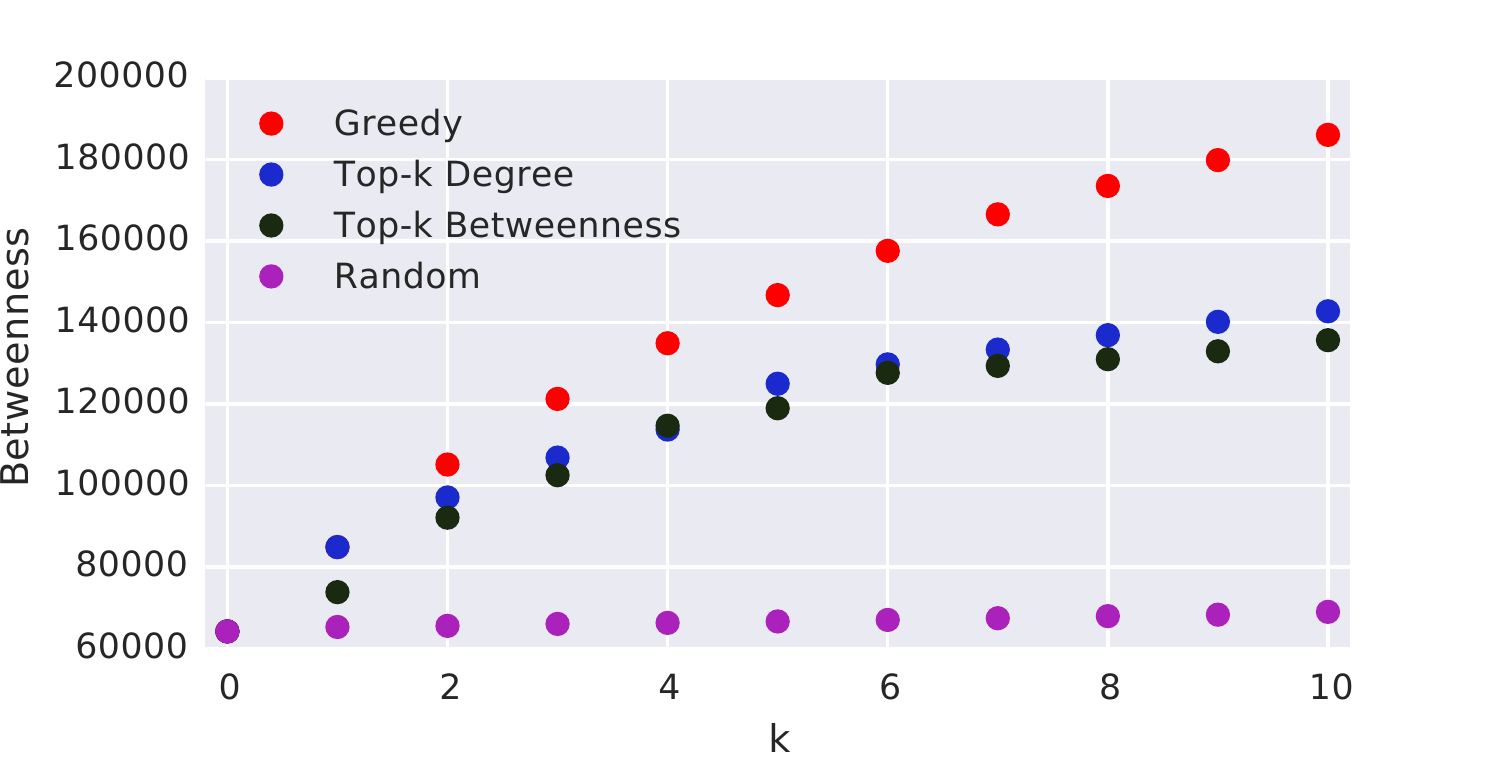}
\caption{Betweenness of the pivot as a function of the number $k$ of inserted edges for the four heuristics. The plots refer to two different pivots in the \texttt{linux} graph.}
\label{fig:bc_linux}
\end{center}
\end{figure}
\begin{figure}[htb]
\begin{center}
\includegraphics[width = 0.49\textwidth]{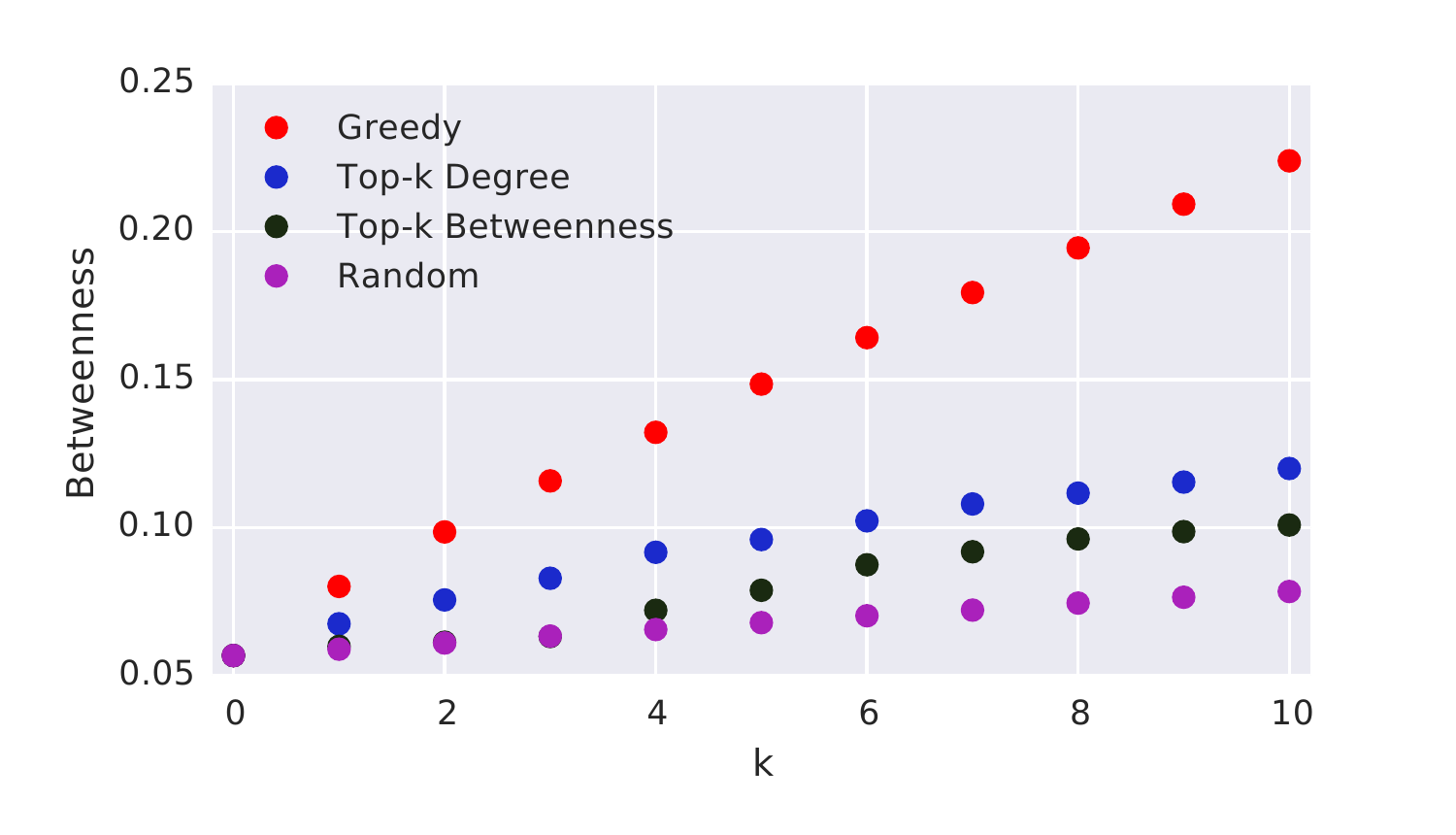}
\includegraphics[width = 0.49\textwidth]{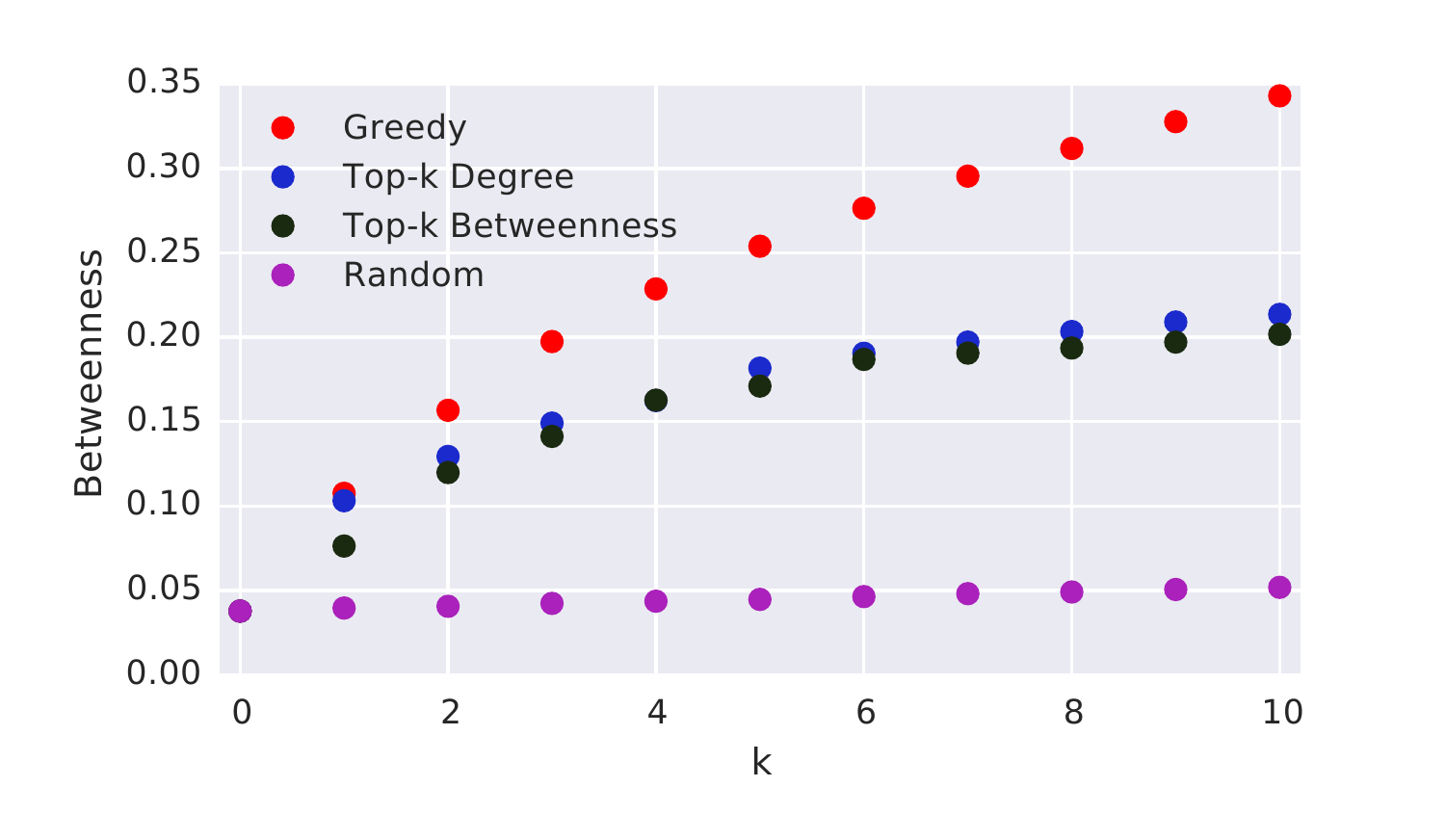}
\caption{Percentage betweenness of the pivot as a function of the number $k$ of inserted edges for the four heuristics. Left: average results for the \texttt{munmun-digg-reply} graph. Right: average results for the \texttt{linux} graph.}
\label{fig:bc_directed}
\end{center}
\end{figure}
\begin{figure}[htb]
\begin{center}
\includegraphics[width = 0.49\textwidth]{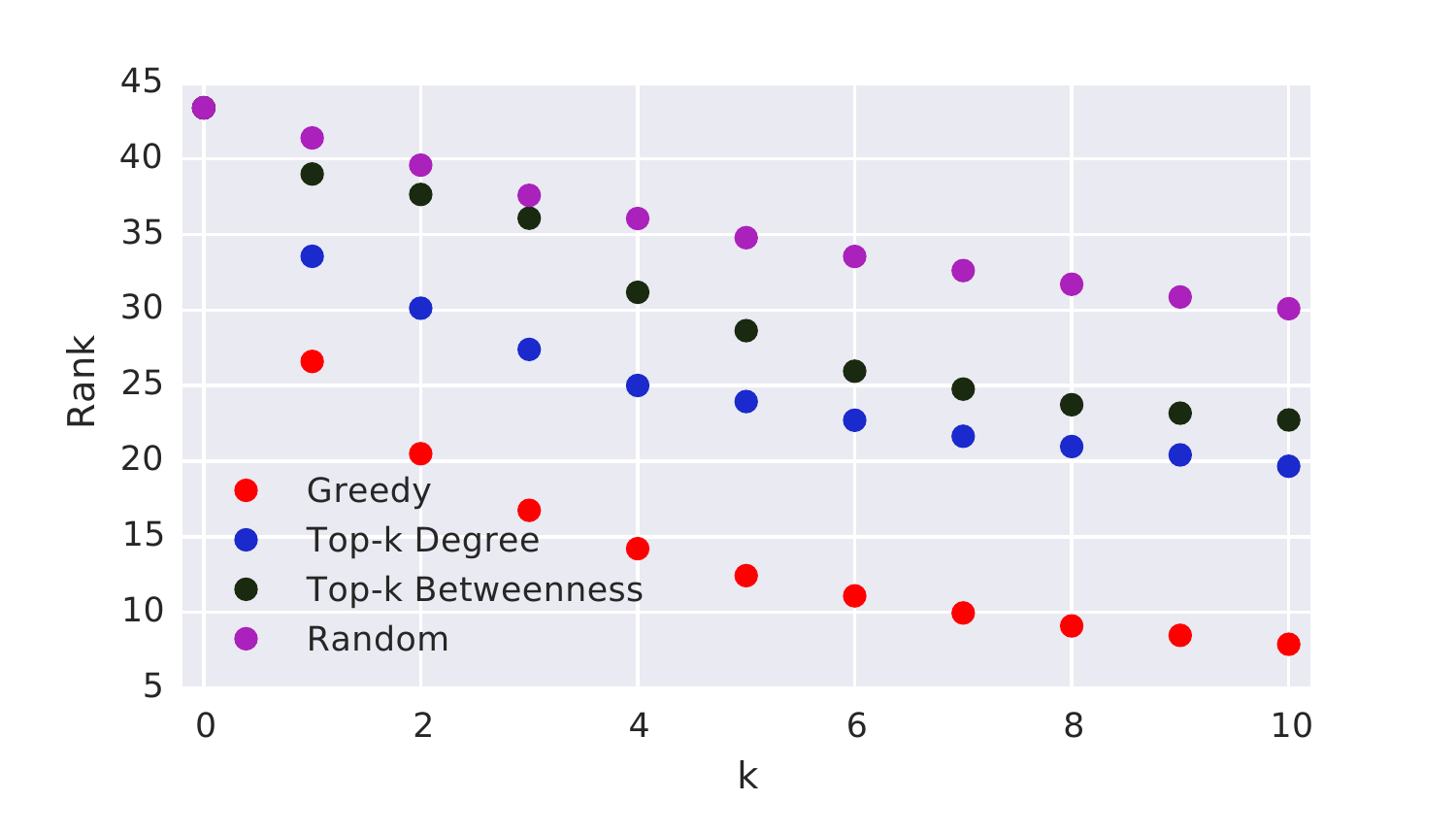}
\includegraphics[width = 0.49\textwidth]{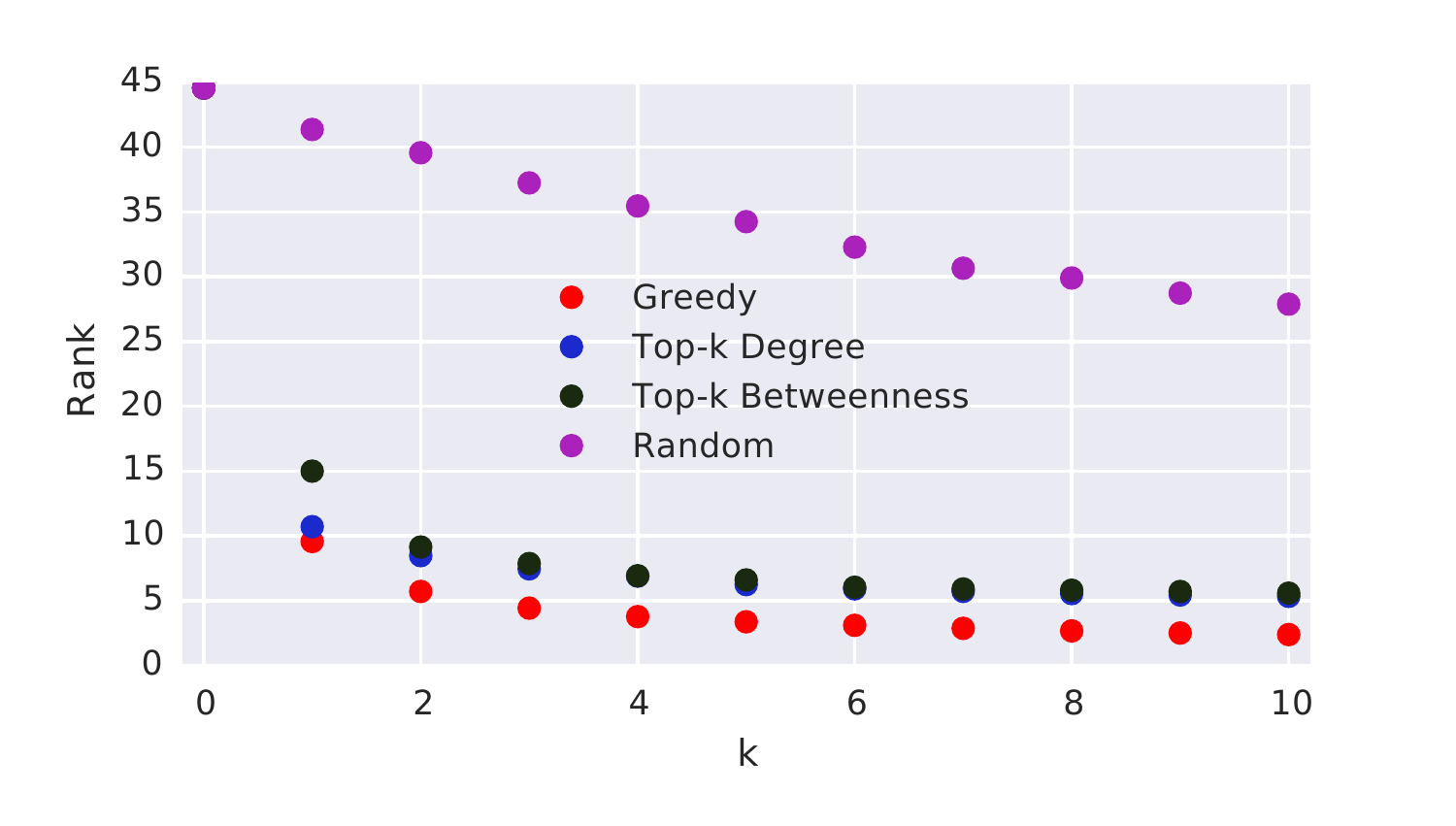}
\caption{Percentage rank of the pivot as a function of the number $k$ of inserted edges for the four heuristics. Left: average results for the \texttt{munmun-digg-reply} graph. Right: average results for the \texttt{linux} graph.}
\label{fig:rank_directed}
\end{center}
\end{figure}
We also analyze the performance of \Greedy on the real-world directed networks of Table~\ref{table:directed} (Section~\ref{sec:experiments}). Since finding the optimum on these networks would take too long, we compare the solution of \Greedy with the following three baselines:
\begin{itemize}
\item \TopKbet \textsc{Degree}: the algorithm that connects the $k$ nodes having the highest degree to $v$;
\item\TopKbet \textsc{Betweenness}: the algorithm that connects the $k$ nodes having the highest betweenness centrality to $v$;
\item \textsc{Random}: the algorithm that connects $k$ nodes extracted uniformly at random to $v$.
\end{itemize}

For each graph, we pick one node at random, compute its betweenness on the initial graph and try to increase it with the four heuristics. We refer to the selected node as \textit{pivot}. Since the results may vary depending on the initial betweenness of the pivot, we also repeat each experiment with 10 different pivots.
In each experiment, we add $k = \{ 1, 2, ..., 10\}$ edges and compute the ranking and betweenness of the pivot after each insertion.

Figure~\ref{fig:bc_munmun} shows the results for the \texttt{munmun-digg-reply} graph, using two different pivots. In particular, the plot on the left shows the betweenness improvement for a node with an initially low betweenness score (close to 0), whereas the one on the right refers to a node that starts with a higher betweenness value (about 40000). Although the final betweenness scores reached by the two nodes differ, we see that the relative performance of the four algorithms is quite similar among the two pivots. A similar behavior can be observed for all other tested pivots. Figure~\ref{fig:bc_linux} reports the results for two different pivots chosen from the \texttt{linux} graph. Again, we notice that the relative performance of the four algorithms is basically the same. Since the same is true also for the other tested graphs, in the following we simply report the average values among the samples.

Figure~\ref{fig:bc_directed} reports the average results (over the sampled pivots) for \texttt{munmun-digg-reply} (left) and \texttt{linux} (right). We define the percentage betweenness of a node $v$ as $b_v \cdot \frac{100}{(n-1)(n-2)}$, where $b_v$ is the betweenness of $v$ and $(n-1)(n-2)$ represents the maximum theoretical betweenness a node can have in a graph with $n$ nodes.
For each value of $k$, the plots show the average percentage betweenness of a pivot after the insertion of $k$ edges (each point represents the average over the 10 pivots). Clearly, the pivot's betweenness after $k$ insertions is a non-decreasing function of $k$, since the insertion of an edge can only increase (or leave unchanged) the betweenness of one of its endpoints.
In both plots, \Greedy outperforms the other heuristics. For example, after 10 edge insertions, the average betweenness of a pivot in the \texttt{munmun-digg-reply} graph is 81460 with \Greedy, 43638 with \TopKbet \textsc{Degree}, 36690 with \TopKbet \textsc{Betweenness} and 28513 with \textsc{Random}. 
A similar behavior can be observed for the average ranks of the pivots, reported in Figure~\ref{fig:rank_directed}. The figures report the percentage ranks, i.e. the ranks multiplied by $\frac{100}{n}$, since $n$ is the maximum rank a node can have in a graph with $n$ nodes. This can be seen as the fraction of nodes with higher betweenness than the pivot.
On \texttt{munmun-digg-reply} , the average initial rank is 2620 (about 43\%). After 10 insertions, the rank obtained using \Greedy is 476 (about 7\%), whereas the one obtained by the other approaches is never lower than 1188 (about 19\%). It is interesting to notice that 3 edge insertions with \Greedy yield a rank of 1011, which is better than the one obtained by the other approaches after 10 insertions. Similarly, also on the \texttt{linux} graph, 3 iterations of \Greedy are enough to bring down the rank from 2498 (45.6\%) to 247 (4.4\%), whereas the other approaches cannot go below 299 (5.3\%) with 10 iterations.
\begin{figure}[tb]
\begin{center}
\includegraphics[width = 0.49\textwidth]{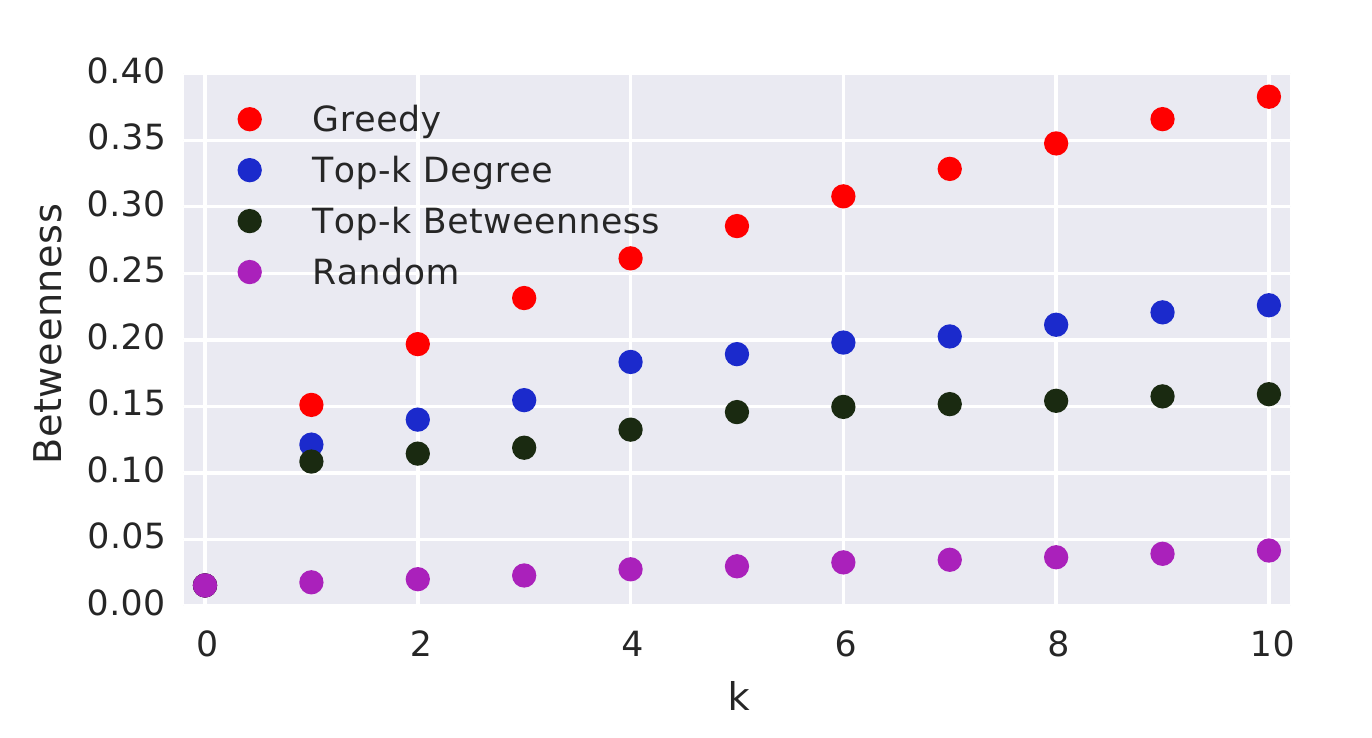}
\includegraphics[width = 0.49\textwidth]{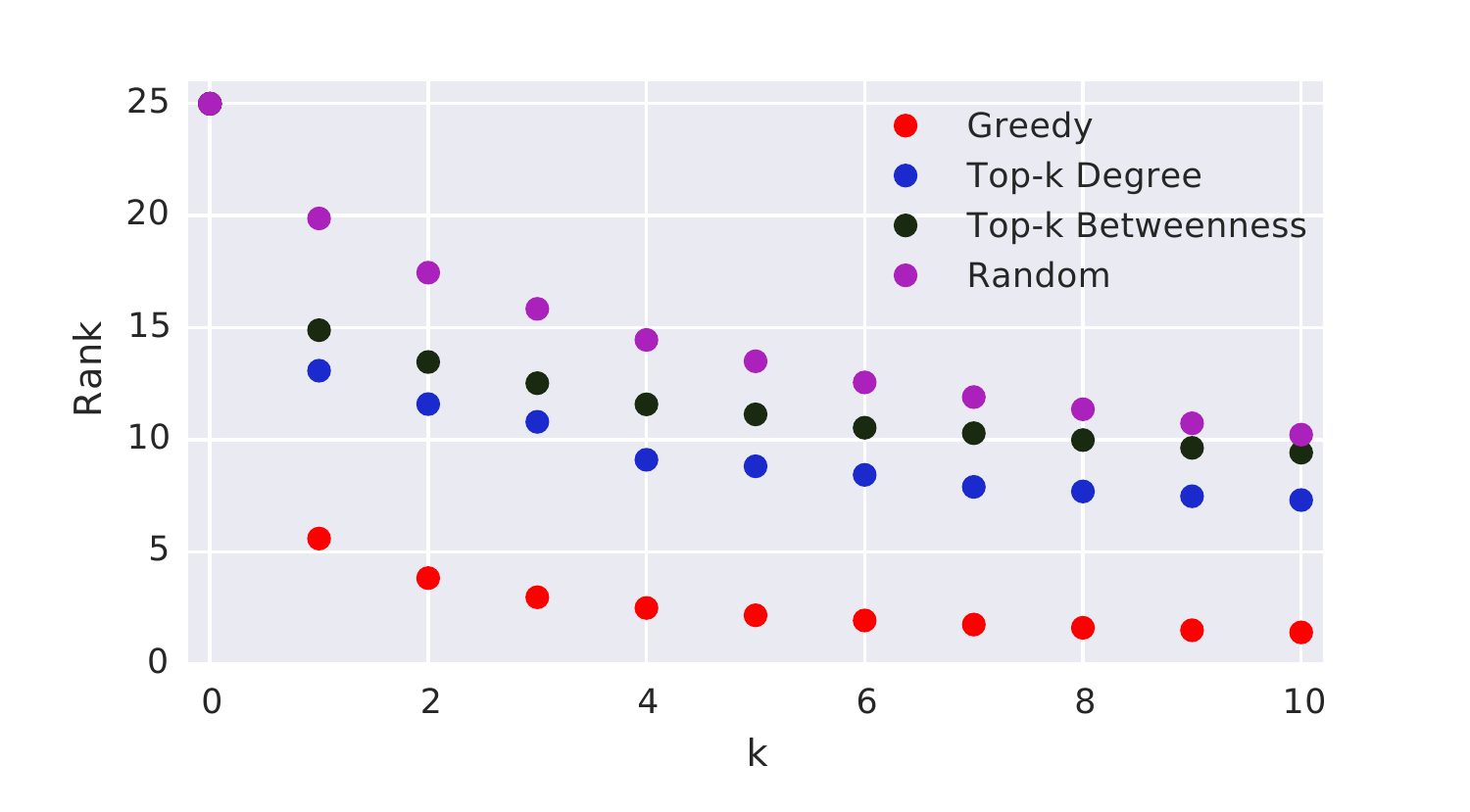}
\caption{Average results over all directed networks. On the left, average percentage betweenness of the pivots as a function of $k$. On the right, average percentage rank of the pivots.}
\label{fig:directed_all}
\end{center}
\end{figure}
Similar results can be observed on the other tested (directed) instances. Figure~\ref{fig:directed_all} reports the average results over all directed networks, both in terms of betweenness (left) and rank (right) improvement. The initial average betweenness of the sample pivots is 0.015\%. \Greedy is by far the best approach, with an average final percentage betweenness (after 10 iterations) of 0.38\% and an average final percentage rank of 1.4\%. As a comparison, the best alternative approach (\TopKbet \textsc{Degree}) yields a percentage betweenness of 0.22\% and a percentage rank of 7.3\%. Not surprisingly, the worst approach is \textsc{Random}, which in 10 iterations yields a final percentage betweenness of 0.04\% and an average percentage rank of 10.2\%.
On average, a single iteration of \Greedy is sufficient for a percentage rank of 5.5\%, better than the one obtained by all other approaches in 10 iterations.
Also, it is interesting to notice that in our experiments \TopKbet \textsc{Degree} performs significantly better than \TopKbet \textsc{Betweenness}. This means that, for the betweenness of a node in a directed graph, it is more important to have incoming edges from nodes with high out-degree than with high betweenness. 
We will see in the following that our results show a different behavior for undirected graphs. 

Also, notice that, although the percentage betweenness scores are quite low, the improvement using \Greedy is still large: with 10 insertions, on average the scores change from an initial 0.015\% to 0.38\%, which is about 25 times the initial value.

\subsubsection{Results for real-world undirected graphs}
Although it was proven that \Greedy has an unbounded approximation ratio for undirected graphs~\cite{DSV15}, it is still not clear how it actually performs in practice.  In~\cite{DSV15}, the authors performed some preliminary experiments in which they showed that the greedy algorithm provides a solution slightly better than the \TopKbet \textsc{Betweenness} algorithm. However, they analyzed only very small networks (with few hundreds of nodes), due to the high complexity of a straightforward implementation of the \Greedy algorithm. In what follows, we compare \Greedy with other baselines on graphs with up to $10^4$ nodes and $10^5$ edges. This is made possible by combining \Greedy with the dynamic algorithm described in Section~\ref{sec:dynamic-single} (note that using the dynamic algorithm only influences the running times of \Greedy, but not its results).
In particular, we compare \Greedy with \TopKbet \textsc{Betweenness}, \TopKbet \textsc{Degree} and \textsc{Random} also on several undirected real-world networks, listed in Table~\ref{table:undirected} of Section~\ref{sec:experiments}.
\begin{figure}[htb]
\begin{center}
\includegraphics[width = 0.49\textwidth]{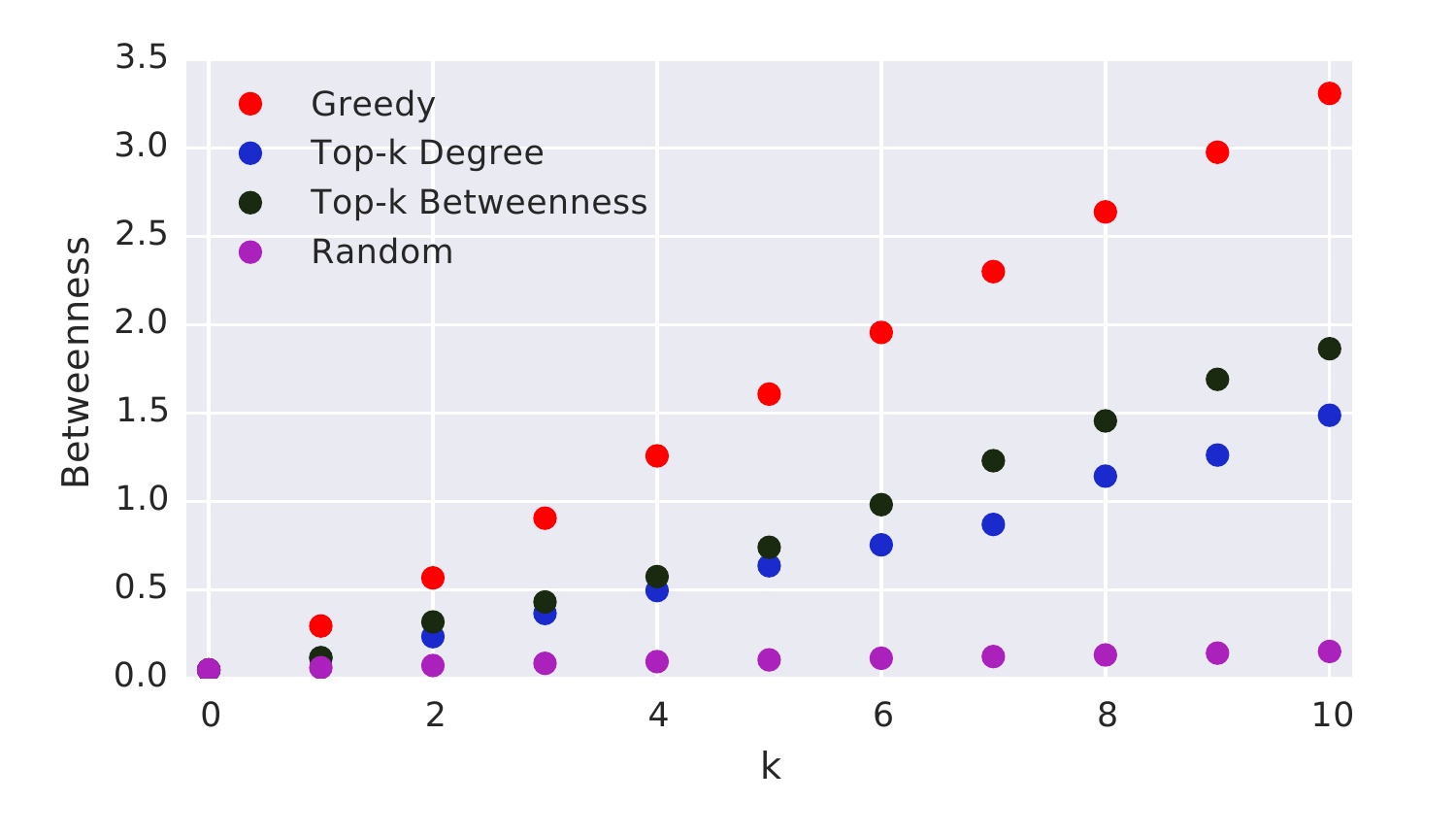}
\includegraphics[width = 0.49\textwidth]{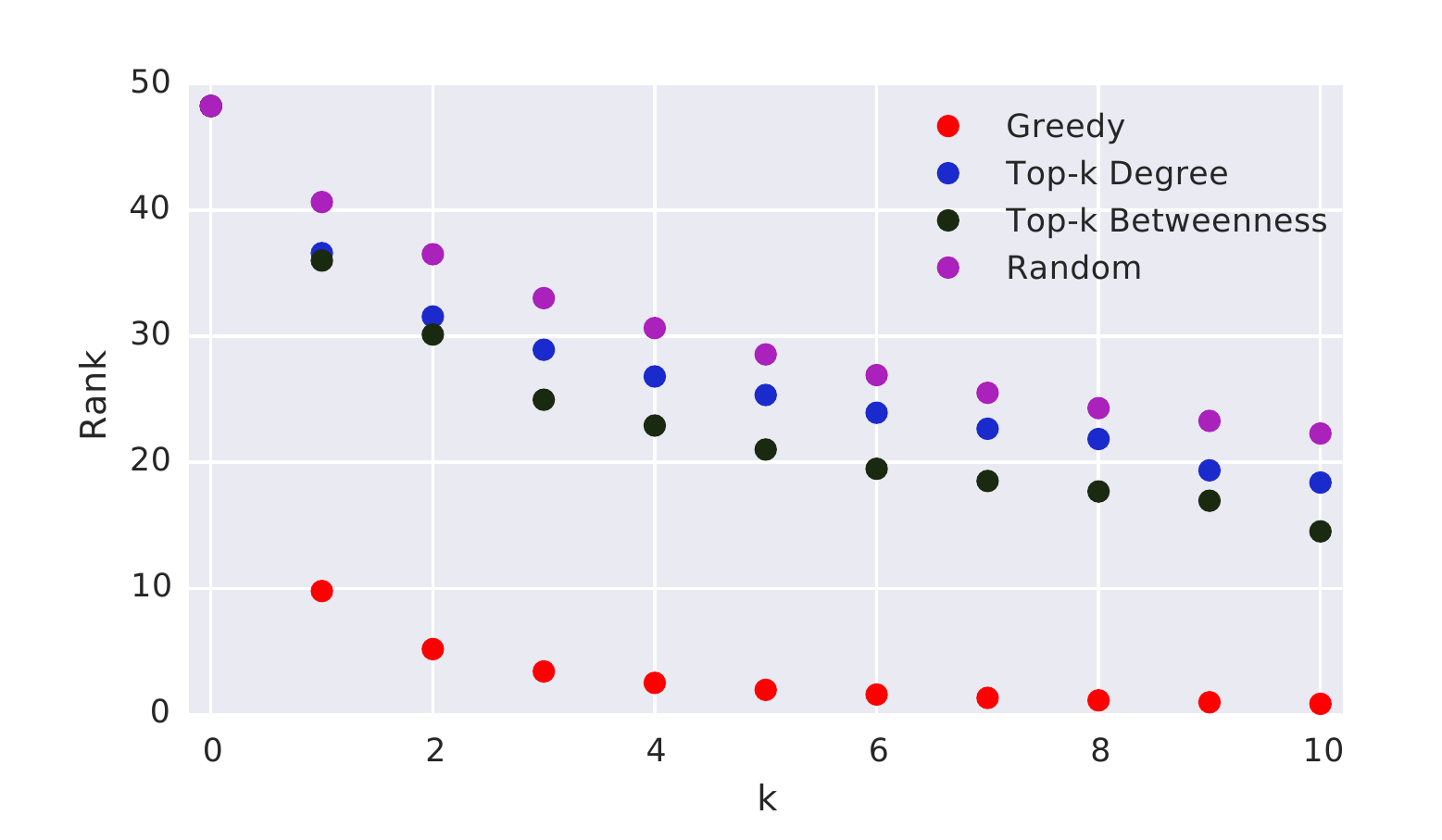}
\caption{Average results over all undirected networks. On the left, average percentage betweenness of the pivots as a function of $k$. On the right, average percentage rank of the pivots.}
\label{fig:undirected_all}
\end{center}
\end{figure}
Figure~\ref{fig:undirected_all} shows the percentage betweenness and ranking, averaged over the undirected networks of Table~\ref{table:undirected}. Also in this case, \Greedy outperforms the other heuristics. In particular, the average initial betweenness of the pivots in the different graphs is 0.05\%. After 10 iterations, the betweenness goes up to 3.7\% with \Greedy, 1.6\% with \TopKbet \textsc{Degree}, 2.1\% with \TopKbet \textsc{Betweenness} and only 0.17\% with \textsc{Random}. The average initial rank is 45\%. \Greedy brings it down to 0.7\% with ten iterations and below 5\% already with two. Using the other approaches, the average rank is always worse than 10\% for \TopKbet \textsc{Betweenness}, 15\% for \TopKbet \textsc{Degree} and 20\% for \textsc{Random}. As mentioned before, differently from directed graphs, \TopKbet \textsc{Betweenness} performs significantly better than \TopKbet \textsc{Degree} in undirected graphs. 

Also, notice that in undirected graphs the percentage betweenness scores of the nodes in the examined graphs are significantly larger than those in the directed graphs. This could be due to the fact that many node pairs have an infinite distance in the examined directed graphs, meaning that these pairs do not contribute to the betweenness of any node. Also, say we want to increase the betweenness of $x$ by adding edge $(v,x)$. The pairs $(s,t)$ for which we can have a shortcut (leading to an increase in the betweenness of $x$) are limited to the ones such that $s$ can reach $v$ and such that $t$ is reachable from $x$, which might be a small fraction of the total number of pairs. On the contrary, most undirected graphs have a giant connected component containing the greatest majority of the nodes. Therefore, it is very likely that a pivot belongs to the giant component or that it will after the first edge insertion.

It is interesting to notice that, despite the unbounded approximation ratio, the improvement achieved by \Greedy on undirected graphs is even larger than for the directed ones: on average 74 times the initial score.
\subsection{Running time evaluation}
\label{sec:experiments}
In this section we evaluate the running time of the dynamic algorithm for betweenness centrality computation. We used the same experimental setting used in Section~\ref{sec:experiments_greedy}. Since some of the algorithms we use for comparison work only on unweighted graphs, all the tested networks are unweighted (although we recall that our algorithm described in Section~\ref{sec:dynamic} can handle also weighted graphs).

\subsubsection{Evaluation of the dynamic algorithm for the betweenness of one node}
\label{sec:experiments_dyn}
In the following, we refer to our incremental algorithm for the update of the betweenness of a single node as \textsf{SI} (Single-node Incremental). Since there are no other algorithms specifically designed to compute or update the betweenness of a single node, we also use the static algorithm by Brandes~\cite{Brandes01betweennessCentrality} and the dynamic algorithm by Bergamini \etal~\cite{DBLP:conf/sea/BergaminiMOS17} for a comparison (the one by Brandes was already in NetworKit). Indeed, the algorithm by Brandes (which we refer to as \textsf{Stat}, from Static) is the best known algorithm for static computation of betweenness and the one by Bergamini \etal (which we name \textsf{AI}, from All-nodes Incremental) has been shown to outperform other dynamic algorithms~\cite{DBLP:conf/sea/BergaminiMOS17}. 

To compare the running times of the algorithms for betweenness centrality, we choose a node $x$ at random and we assume we want to compute the betweenness of $x$. Then, we add an edge to the graph, also chosen uniformly at random among the node pairs $(u,v)$ such that $(u,v) \notin E$. After the insertion, we use the three algorithms to update the betweenness centrality of $x$ and compare their running times. We recall that \textsf{Stat} is a static algorithm, which means that we can only run it from scratch on the graph after the edge insertion. On each graph, we repeat this 100 times and report the average running time obtained by each of the algorithms.

\begin{table}[tb]
\caption{Average running times of the betweenness algorithms on directed real-world graphs. The last two columns report the standard deviation of the running times of \textsf{AI} and \textsf{SI} over the 100 edge insertions.}{
\begin{scriptsize}
  \begin{tabular}{ | l | r | r | r | r | r | r | r |}
    \hline
Graph	&	Nodes	&	Edges	&	Time \textsf{Stat} [s]	&	Time \textsf{AI} [s]	&	Time \textsf{SI} [s]	&	STD \textsf{AI} [s] &  STD {SI} [s]\\
\hline
\texttt{subelj-jung}	&	6 120	&	50 535	& 1.25	 & 	0.0019	 & 	 \textbf{0.0002}	 & 	0.0036	 & 	0.0005 \\
\texttt{wiki-Vote}	&	7 115	&	100 762	& 8.18	 & 	0.0529	 & 	 \textbf{0.0015}	 & 	0.0635	 & 	0.0038 \\
\texttt{elec}	&	7 118	&	103 617	& 8.67	 & 	0.0615	 & 	 \textbf{0.0019}	 & 	0.0858	 & 	0.0053 \\
\texttt{freeassoc}	&	10 617	&	63 788	& 14.96 & 	0.1118	 & 	 \textbf{0.0034}	 & 	0.1532	 & 	0.0036 \\
\texttt{dblp-cite}	&	12 591	&	49 728	& 5.04	 & 	0.1726	 & 	 \textbf{0.0071}	 & 	0.7905	 & 	0.0451 \\
\texttt{subelj-cora}	&	23 166	&	91 500	& 34.08	 & 	0.3026	 & 	 \textbf{0.0327}	 & 	1.1598	 & 	0.1575 \\

\texttt{ego-twitter}	&	23 370	&	33 101	& 8.47	 & 	0.0062	 & 	 \textbf{0.0001}	 & 	0.0576	 & 	0.0003 \\
\texttt{ego-gplus}	&	23 628	&	39 242	& 10.01	 & 	0.0024	 & 	 \textbf{0.0001}	 & 	0.0026	 & 	0.0000 \\
\texttt{munmun-digg}	&	30 398	&	85 247	& 78.09	 & 	0.2703	 & 	 \textbf{0.0073}	 & 	0.2539	 & 	0.0099 \\
\texttt{linux}	&	30 837	&	213 424	& 34.75	 & 	0.0692	 & 	 \textbf{0.0108}	 & 	0.3019	 & 	0.0637 \\
\hline
  \end{tabular}
  \end{scriptsize}
}
 \label{table:directed}
\end{table}
\begin{figure}[tb]
\begin{center}
\includegraphics[width = 0.49\textwidth]{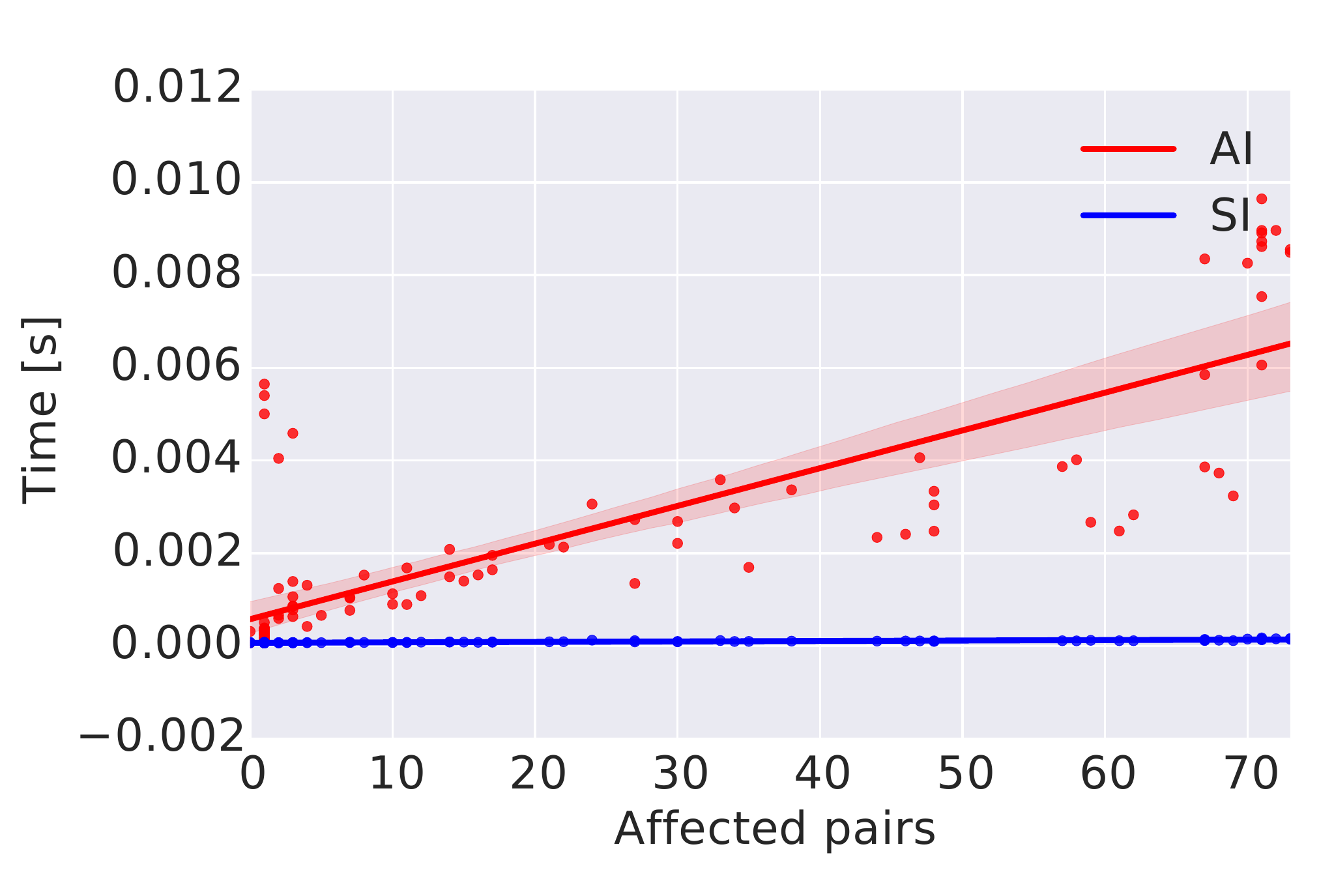}
\includegraphics[width = 0.49\textwidth]{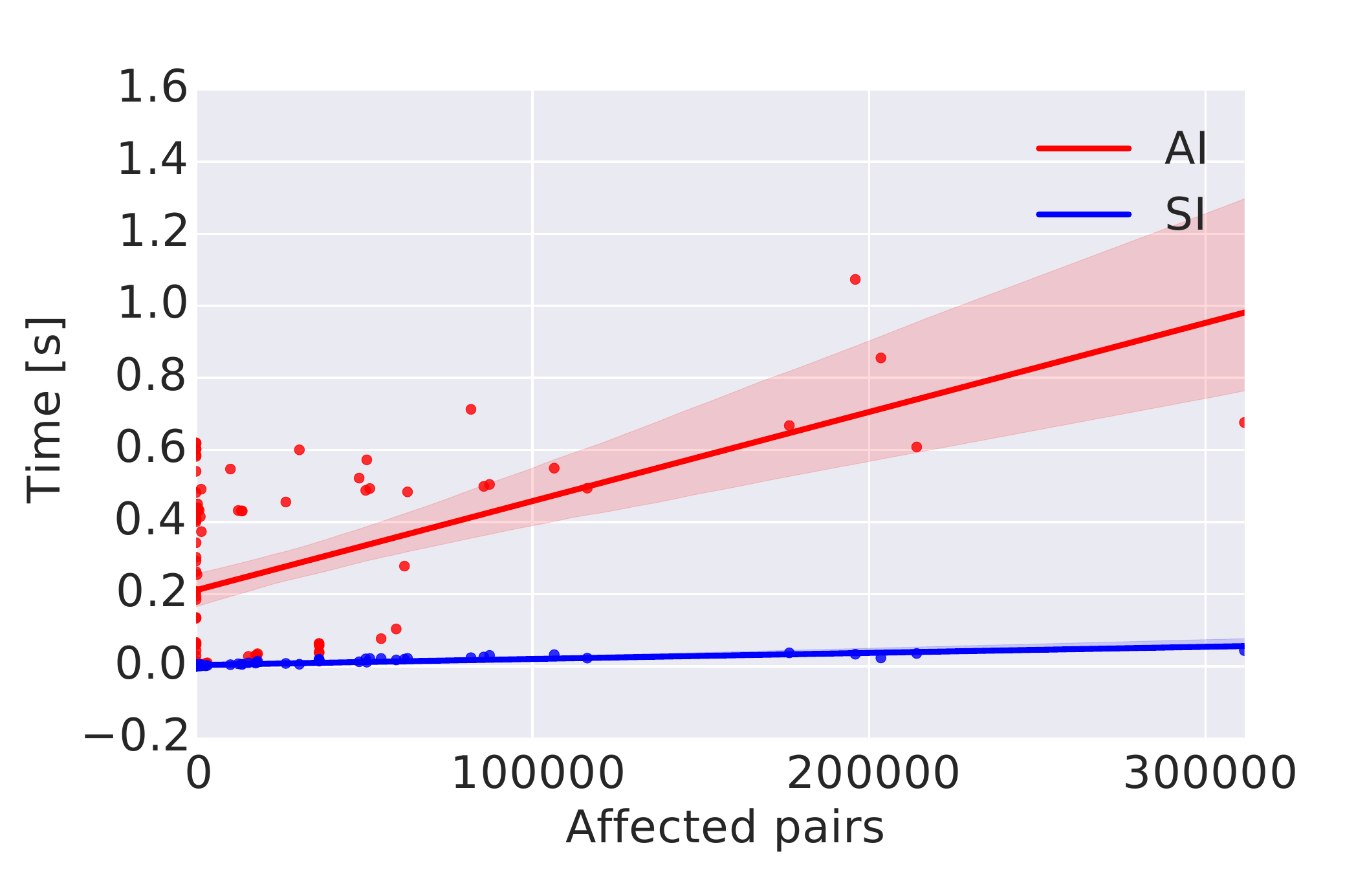}
\includegraphics[width = 0.49\textwidth]{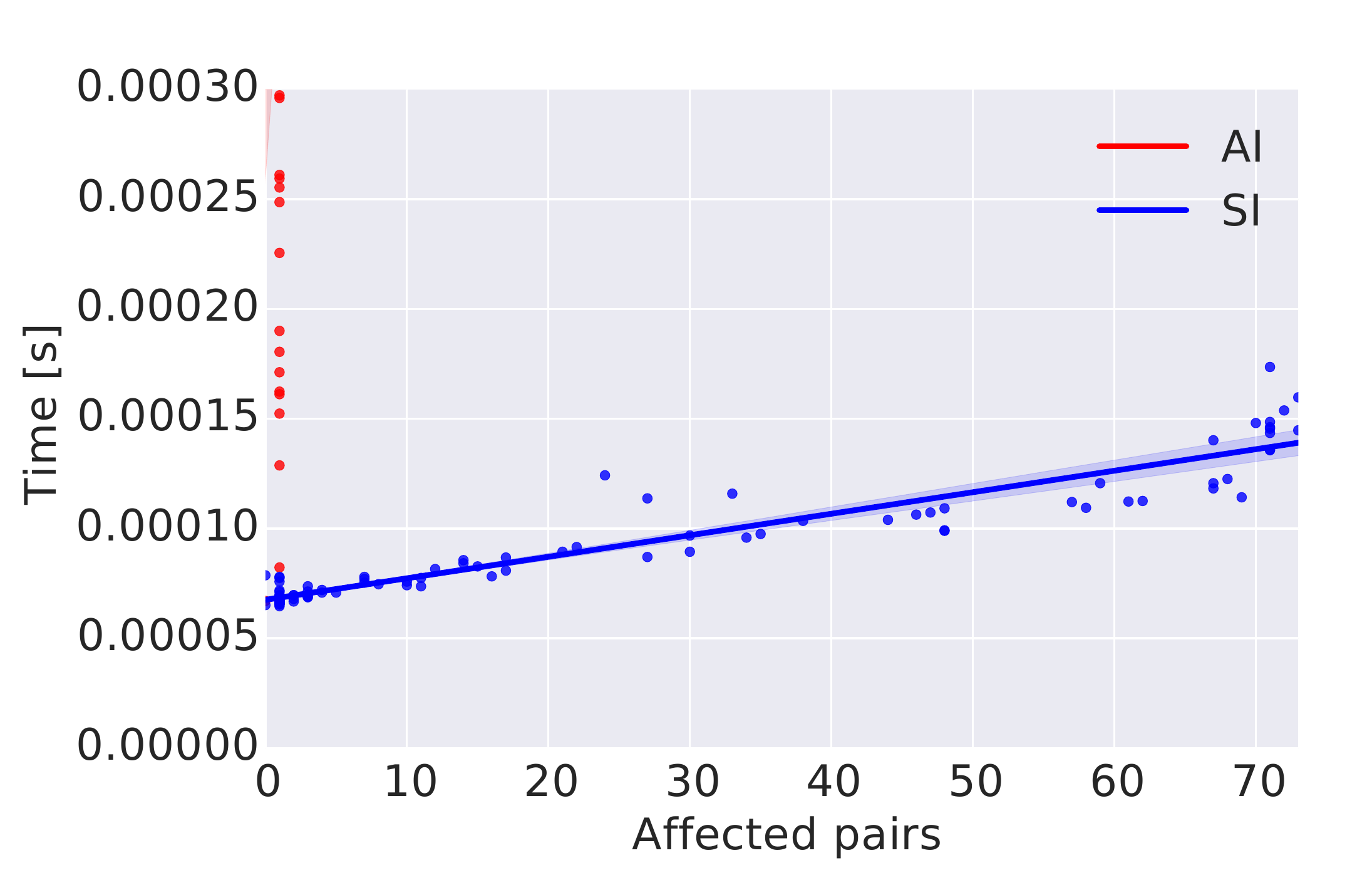}
\includegraphics[width = 0.49\textwidth]{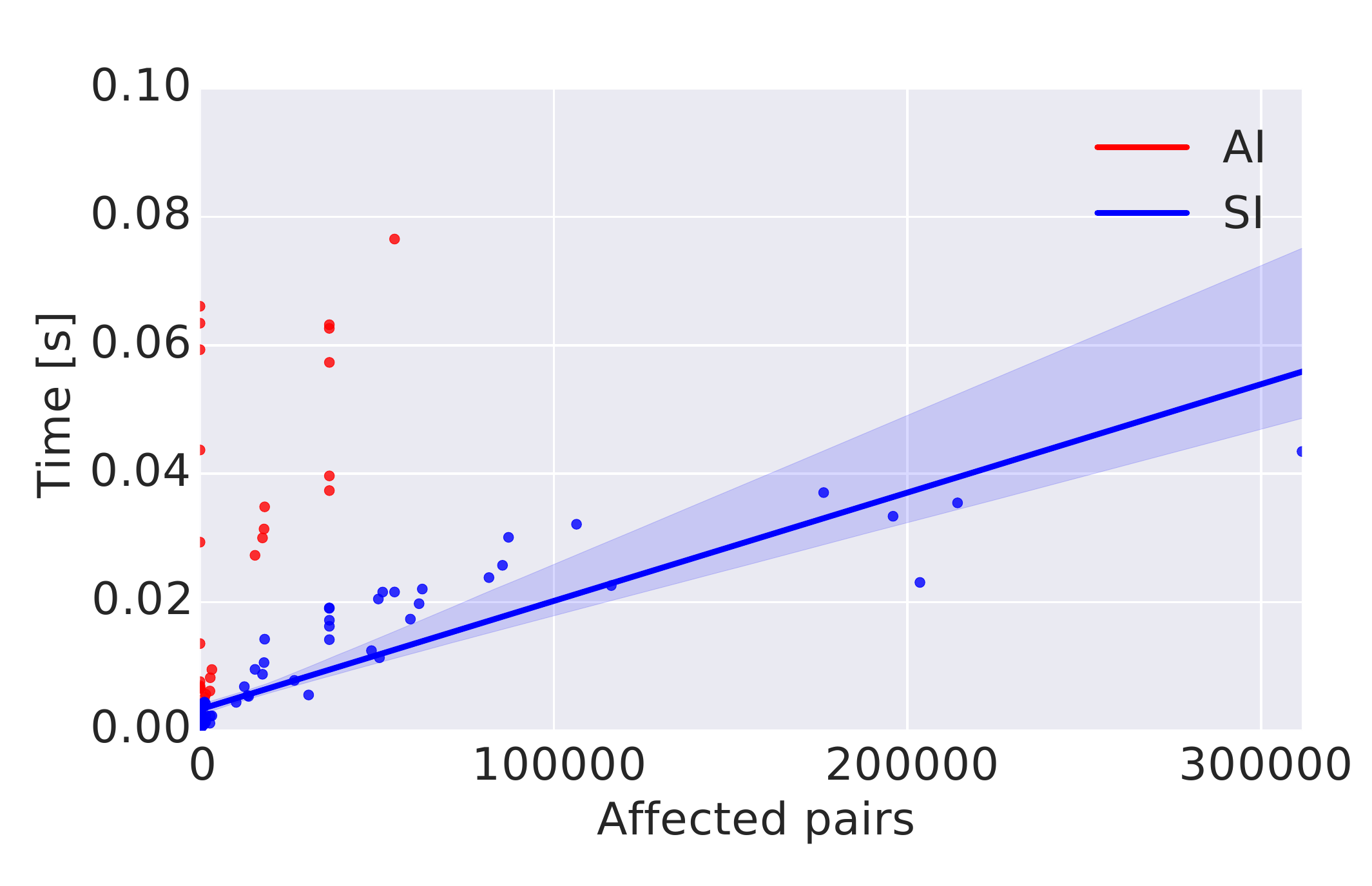}
\caption{Top: Running time of \textsf{AI} and \textsf{SI} as a function of the number of affected node pairs for two directed graphs (left: \texttt{ego-gplus}, right: \texttt{munmun-digg}). Bottom: Same as the two plots above, but zoomed on the running times of \textsf{SI}. The points are the computed running times, the lines are the results of a linear regression and the area around the lines is a 95\% confidence interval for the regression.}
\label{fig:aff_directed}
\end{center}
\end{figure}
\begin{table}[tb]
\caption{Average running times of the betweenness algorithms on undirected real-world graphs. The last two columns report the standard deviation of the running times of \textsf{AI} and \textsf{SI} over the 100 edge insertions.}{
\begin{scriptsize}
  \begin{tabular}{ | l | r | r | r | r | r | r | r |}
    \hline
Graph	&	Nodes	&	Edges	&	Time \textsf{Stat} [s]	&	Time \textsf{AI} [s]	&	Time \textsf{SI} [s]	&	SD \textsf{AI} [s] &  SD {SI} [s]\\

\hline
\texttt{Mus-musculus}	&	4 610	&	5 747	& 2.87	 & 	0.0337	 & 	 \textbf{0.0037}	 & 	0.0261	 & 	0.0024 \\
\texttt{HC-BIOGRID}	&	4 039	&	10 321	& 5.32	 & 	0.1400	 & 	 \textbf{0.0083}	 & 	0.1450	 & 	0.0119 \\
\texttt{Caenor-eleg}	&	4 723	&	9 842	& 4.75	 & 	0.0506	 & 	 \textbf{0.0025}	 & 	0.0406	 & 	0.0014 \\
\texttt{ca-GrQc}	&	5 241	&	14 484	& 4.15	 & 	0.0377	 & 	 \textbf{0.0033}	 & 	0.0245	 & 	0.0017 \\
\texttt{advogato}	&	7 418	&	42 892	& 12.65	 & 	0.1820	 & 	 \textbf{0.0024}	 & 	0.1549	 & 	0.0008 \\
\texttt{hprd-pp}	&	9 465	&	37 039	& 29.19	 & 	0.2674	 & 	 \textbf{0.0053}	 & 	0.1873	 & 	0.0021 \\
\texttt{ca-HepTh}	&	9 877	&	25 973	& 21.57	 & 	0.1404	 & 	 \textbf{0.0095}	 & 	0.1108	 & 	0.0053 \\
\texttt{dr-melanog}	&	10 625	&	40 781	&38.18	 & 	0.2687	 & 	 \textbf{0.0067}	 & 	0.2212	 & 	0.0029 \\

\texttt{oregon1}	&	11 174	&	23 409	& 23.77	 & 	0.5676	 & 	 \textbf{0.0037}	 & 	0.5197	 & 	0.0020 \\

\texttt{oregon2}	&	11 461	&	32 730	& 27.98	 & 	0.5655	 & 	 \textbf{0.0039}	 & 	0.5551	 & 	0.0026 \\
\texttt{Homo-sapiens}	&	13 690	&	61 130	& 68.06	 & 	0.5920	 & 	 \textbf{0.0079}	 & 	0.4203	 & 	0.0035 \\
\texttt{GoogleNw}	&	15 763	&	148 585	& 76.17 & 	2.4744	 & 	 \textbf{0.0044}	 & 	4.1075	 & 	0.0045 \\
\texttt{CA-CondMat} & 21 363 	&	91 342 & 168.44	 & 	1.1375	 & 	 \textbf{0.0486}	 & 	0.7485	 & 	0.0358 \\
\hline
  \end{tabular}
  \end{scriptsize}
}
  \label{table:undirected}
\end{table}
\begin{figure}[tb]
\begin{center}
\includegraphics[width = 0.49\textwidth]{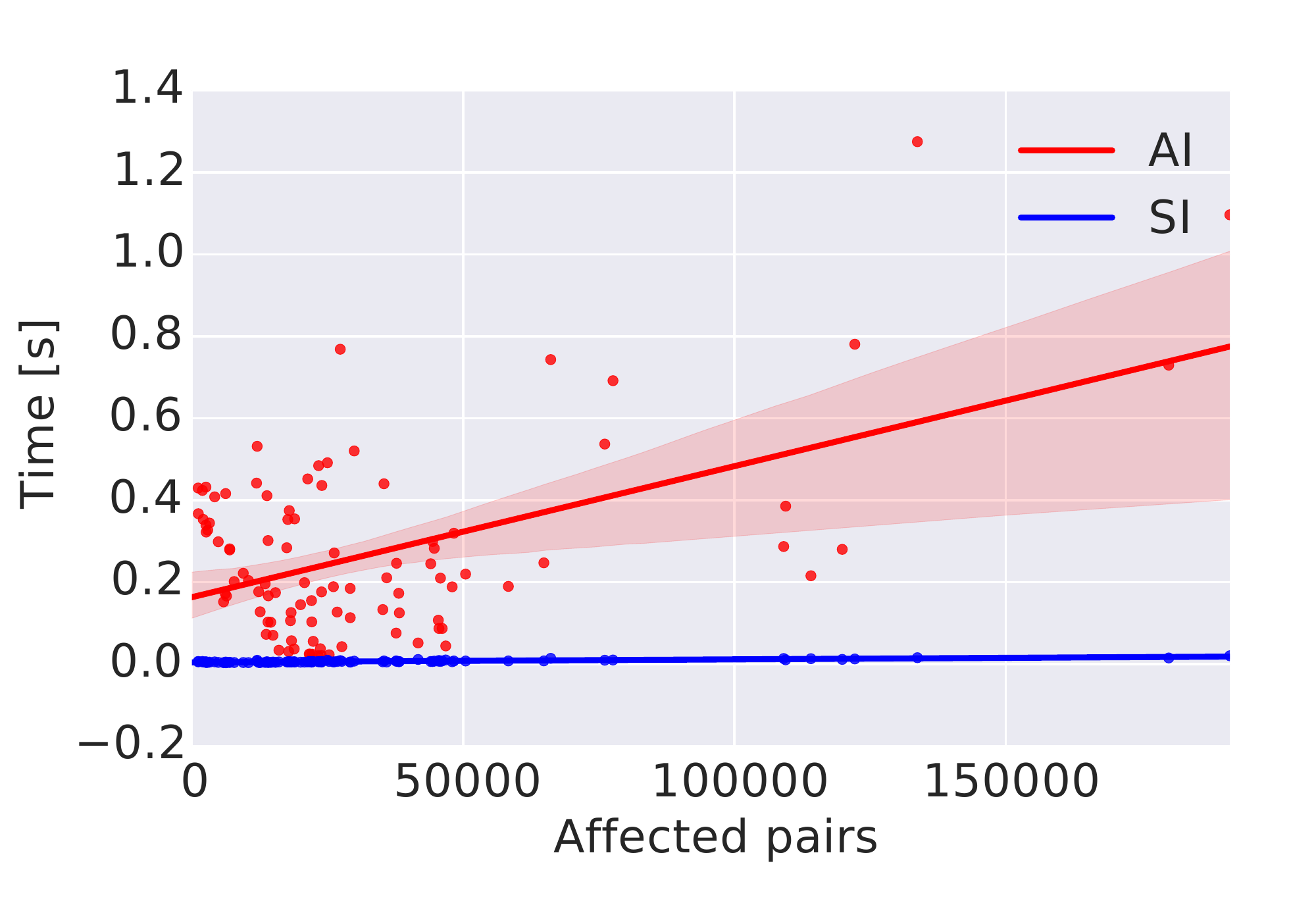}
\includegraphics[width = 0.49\textwidth]{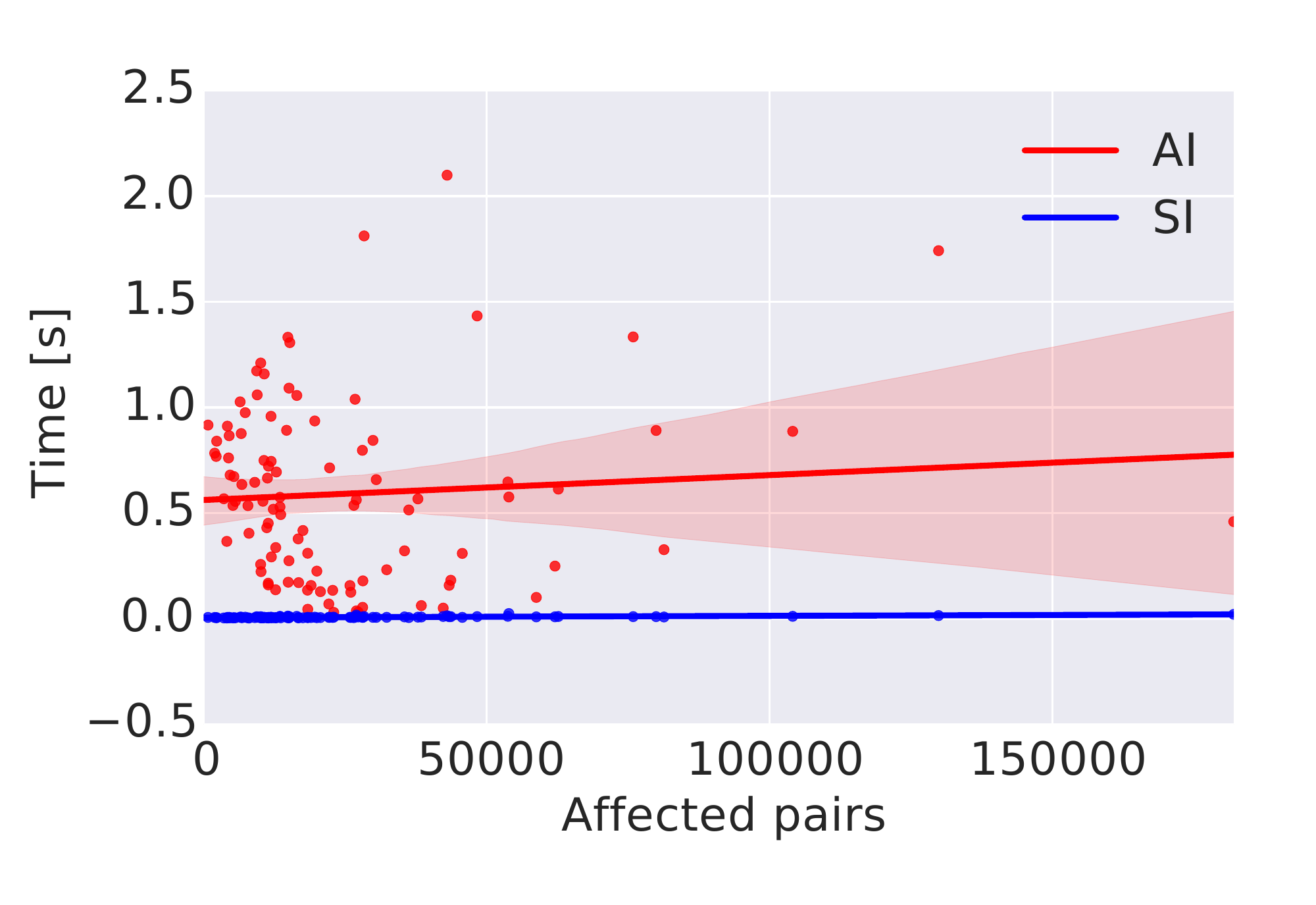}
\includegraphics[width = 0.49\textwidth]{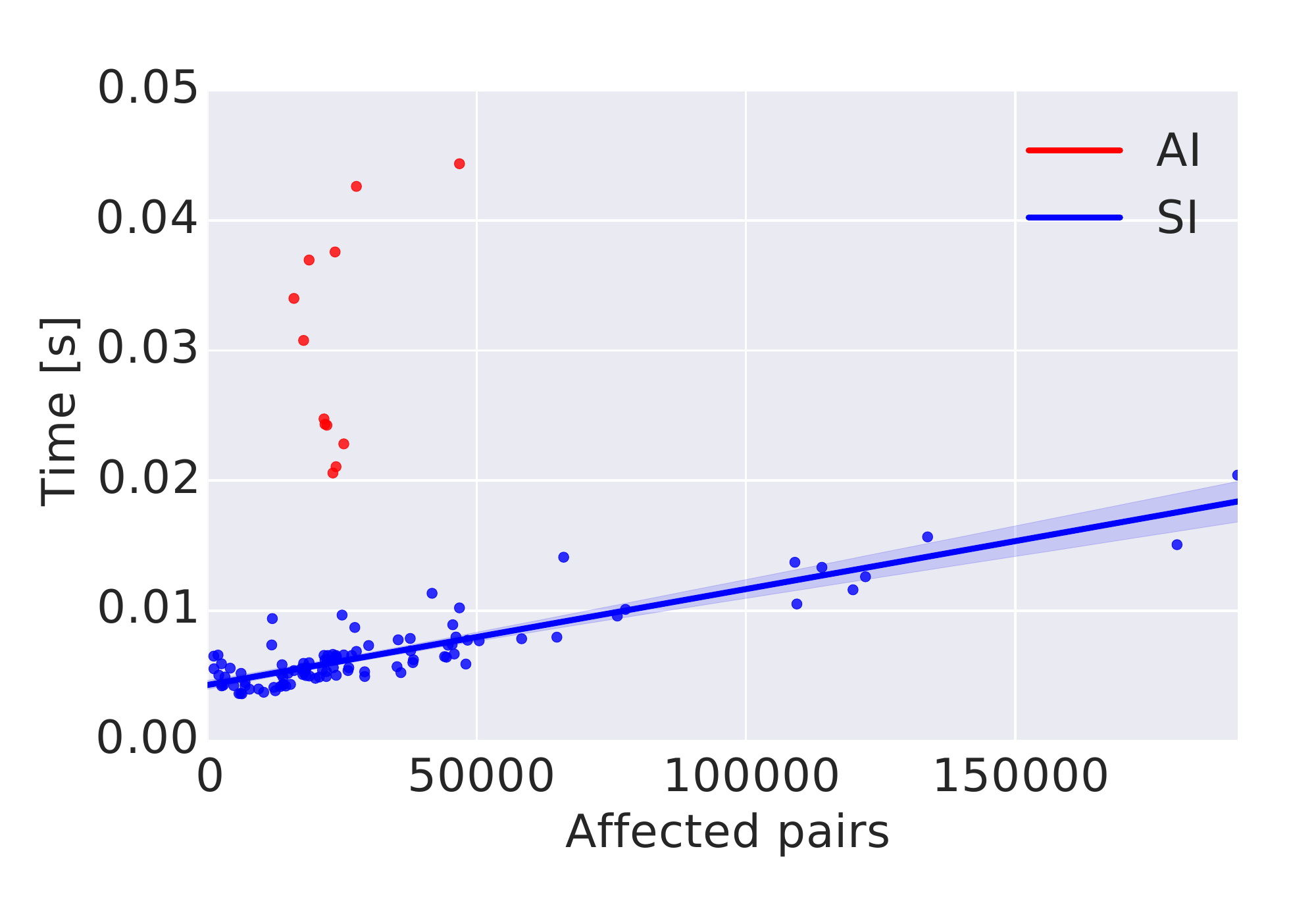}
\includegraphics[width = 0.49\textwidth]{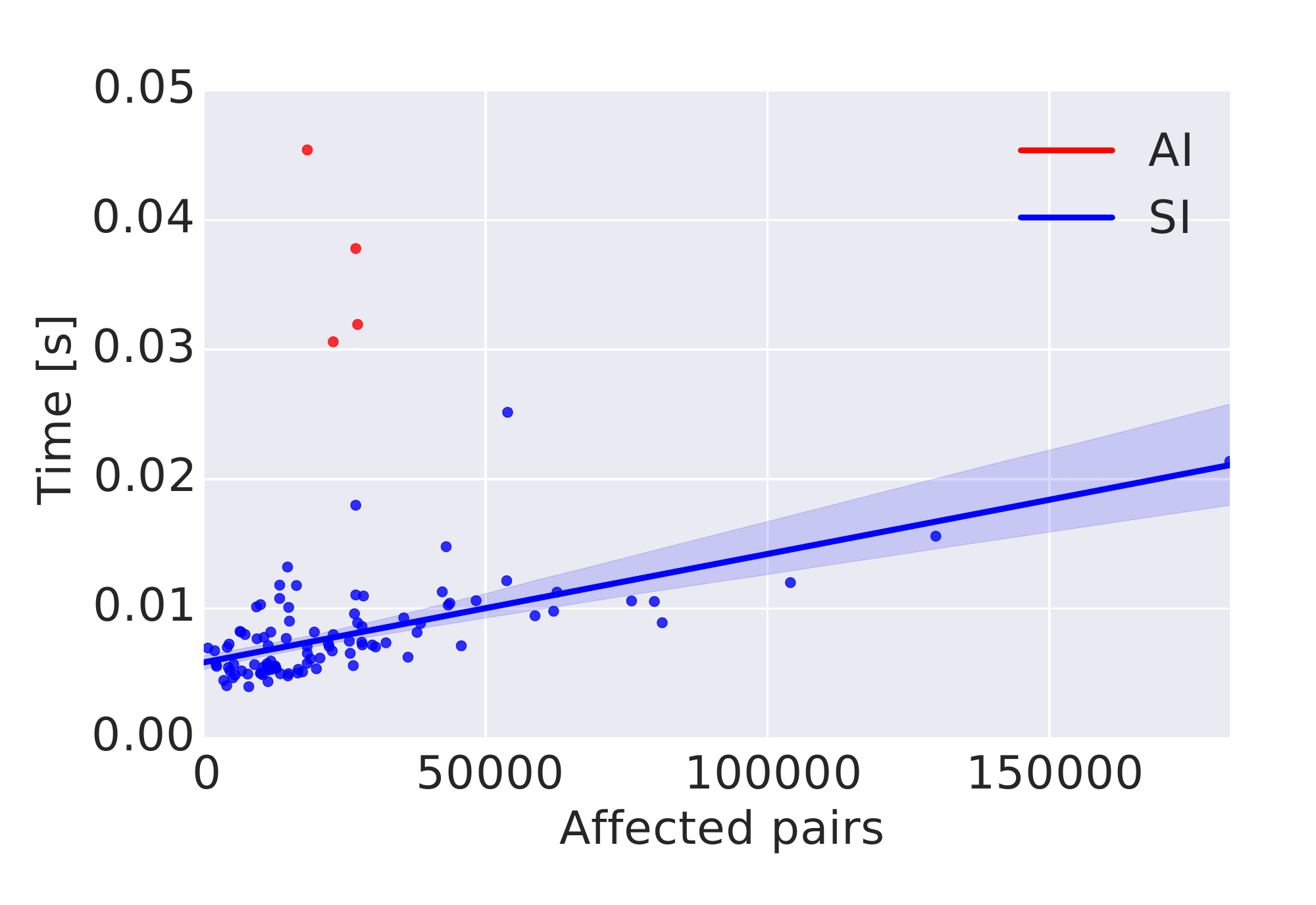}
\caption{Top: Running time of \textsf{AI} and \textsf{SI} as a function of the number of affected node pairs for two undirected graphs (left: \texttt{dr-melanog}, right: \texttt{Homo-sapiens}). Bottom: same as the two plots above, but zoomed on the running times of \textsf{SI}. The points are the computed running times, the lines are the results of a linear regression and the area around the lines is a 95\% confidence interval for the regression.}
\label{fig:aff_undirected}
\end{center}
\end{figure}
Table~\ref{table:directed} and  Table~\ref{table:undirected} show the running times for directed and undirected graphs, respectively. As expected, both dynamic algorithms \textsf{AI} and \textsf{SI} are faster than the static approach and \textsf{SI} is the fastest among all algorithms. This is expected, since \textsf{SI} is the one that performs the smallest number of operations. Also, notice that the standard deviation of the running times of both \textsf{AI} and \textsf{SI} is very high, sometimes even higher than the average. This is actually not surprising, since different edge insertion might affect portions of the graph of very different sizes.
Figure~\ref{fig:aff_directed} and Figure~\ref{fig:aff_undirected} report the running times of  \textsf{AI} and \textsf{SI} as a function of the number of affected node pairs for two directed and undirected graphs, respectively (similar results can be observed for the other tested graphs). As expected, the running time of both algorithms (as well as the difference between the running time of \textsf{AI} and that of \textsf{SI}) mostly increases as the number of affected pairs increases. However, \textsf{AI} presents a much larger deviation than \textsf{SI}. This is due to the fact that its running time also depends on the number of nodes that used to lie in old shortest paths between the affected pairs. Indeed, the number of nodes whose betweenness gets affected does not only depend on the number of affected pairs (which we recall to be the ones for which the edge insertion creates a shortcut or a new shortest paths), but also on how many shortest paths there used to be between the affected pairs before the insertion and how long  these paths were. 
\begin{table}[tb]
\caption{Speedups on the static algorithm and on the dynamic algorithm for all nodes on directed networks. For both \textsf{Stat} and \textsf{AI}, the first column reports the geometric mean of the speedups over the 100 insertions, the second column reports the maximum speedups and the third column the minimum speedup.}{
\begin{scriptsize}
  \begin{tabular}{ | l | r | r | r | r | r | r | }
  \cline{2-7}
  \multicolumn{1}{ l |}{} & \multicolumn{3}{| c |}{Speedups on \textsf{Stat}} & \multicolumn{3}{| c |}{Speedups on \textsf{AI}}\\
    \hline
Graph	&	Geometric mean	&	Maximum	&	Minimum	&	Geometric mean	&	Maximum	&	Minimum \\
\hline
\texttt{subelj-jung}	&	24668.3	 & 	67477.6	 & 	342.9	 & 	10.0	 & 	63.7	 & 	1.1 \\
\texttt{wiki-Vote}	&	23779.8	 & 	381357.7	 & 	275.6	 & 	39.3	 & 	310.7	 & 	1.0 \\
\texttt{elec}	&	21560.5	 & 	408629.3	 & 	175.6	 & 	32.1	 & 	285.1	 & 	1.1 \\
\texttt{freeassoc}	&	6783.2	 & 	330333.4	 & 	707.5	 & 	13.0	 & 	94.0	 & 	1.1 \\
\texttt{dblp-cite}	&	24745.7	 & 	140950.0	 & 	11.4	 & 	13.3	 & 	314.5	 & 	1.1 \\
\texttt{subelj-cora}	&	18936.5	 & 	543630.2	 & 	22.9	 & 	32.4	 & 	257.5	 & 	1.0 \\
\texttt{ego-twitter}	&	111597.2	 & 	134716.0	 & 	3169.2	 & 	4.0	 & 	216.3	 & 	1.0 \\
\texttt{ego-gplus}	&	115936.2	 & 	154869.5	 & 	57650.6	 & 	14.6	 & 	74.5	 & 	1.1 \\
\texttt{munmun-digg}	&	34299.5	 & 	998564.0	 & 	1796.8	 & 	30.7	 & 	188.3	 & 	1.2 \\
\texttt{linux}	&	103469.6	 & 	433745.5	 & 	59.5	 & 	32.1	 & 	94.6	 & 	1.5 \\
\hline
  \end{tabular}
  \end{scriptsize}
}
 \label{table:directed_cont}
\end{table}

\begin{table}[tb]
\caption{Speedups on the static algorithm and on the dynamic algorithm for all nodes on undirected networks. For both \textsf{Stat} and \textsf{AI}, the first column reports the geometric mean of the speedups over the 100 insertions, the second column reports the maximum speedups and the third column the minimum speedup.}{
\begin{scriptsize}
  \begin{tabular}{ | l | r | r | r | r | r | r | }
  \cline{2-7}
  \multicolumn{1}{ l |}{} & \multicolumn{3}{| c |}{Speedups on \textsf{Stat}} & \multicolumn{3}{| c |}{Speedups on \textsf{AI}}\\
    \hline
Graph	&	Geometric mean	&	Maximum	&	Minimum	&	Geometric mean	&	Maximum	&	Minimum \\
\hline
\texttt{Mus-musculus}	&	1031.2	 & 	174166.8	 & 	191.4	 & 	7.7	 & 	21.9	 & 	1.7 \\
\texttt{HC-BIOGRID}	&	962.3	 & 	4060.6	 & 	56.7	 & 	17.0	 & 	51.1	 & 	4.5 \\
\texttt{Caenor-eleg}	&	2152.2	 & 	293172.8	 & 	474.6	 & 	15.0	 & 	49.4	 & 	1.3 \\
\texttt{ca-GrQc}	&	1517.5	 & 	220289.2	 & 	351.8	 & 	10.0	 & 	22.8	 & 	2.1 \\
\texttt{advogato}	&	5819.0	 & 	698406.6	 & 	2860.5	 & 	43.4	 & 	192.7	 & 	1.9 \\
\texttt{hprd-pp}	&	5846.6	 & 	10852.3	 & 	1696.6	 & 	39.2	 & 	119.5	 & 	3.2 \\
\texttt{ca-HepTh}	&	2642.6	 & 	432794.2	 & 	549.2	 & 	12.3	 & 	35.7	 & 	2.8 \\
\texttt{dr-melanog}	&	6105.9	 & 	10589.7	 & 	1869.7	 & 	29.9	 & 	88.3	 & 	3.1 \\
\texttt{oregon1}	&	7407.4	 & 	733008.3	 & 	1562.3	 & 	72.6	 & 	493.4	 & 	2.4 \\
\texttt{oregon2}	&	9192.5	 & 	617710.0	 & 	1595.8	 & 	68.8	 & 	470.5	 & 	2.5 \\
\texttt{Homo-sapiens}	&	9216.4	 & 	17177.4	 & 	2706.2	 & 	57.0	 & 	165.2	 & 	3.4 \\
\texttt{GoogleNw}	&	34967.2	 & 	505509.9	 & 	3799.3	 & 	137.3	 & 	1560.3	 & 	2.9 \\
\texttt{CA-CondMat} & 4073.9	 & 	10690.6	 & 	537.8	 & 	20.8	 & 	69.4	 & 	2.8 \\
\hline
  \end{tabular}
  \end{scriptsize}
}
 \label{table:undirected_cont}
\end{table}
Table~\ref{table:directed_cont} and Table~\ref{table:undirected_cont} show the speedups of \textsf{SI} on \textsf{AI} and those of \textsf{SI} on \textsf{Stat}, for directed and undirected graphs, respectively. 
Although the speedups vary considerably among the networks and the edge insertions, \textsf{SI} is always at least as fast as \textsf{AI} and up to 1560 times faster (maximum speedup for \texttt{GoogleNw}). On average (geometric mean of the average speedups over the tested networks), \textsf{SI} is 29 times faster than \textsf{AI} for undirected graphs and 18 times faster for directed graphs. 
The high speedups on the dynamic algorithm for all nodes is due to the fact that, when focusing on a single node, we do not need to update the scores of all the nodes that lie in some shortest path affected by the edge insertion. 
On the contrary, for each affected source node $s$, \textsf{AI} has to recompute the change in dependencies by iterating over all nodes that lie in either a new or an old shortest path from $s$.
As a result, \textsf{SI} is extremely fast: on all tested instances, its running time is always smaller than 0.05 seconds, whereas \textsf{AI} can take up to seconds to update betweenness.

Compared to recomputation, \textsf{SI} is on average about 4200 times faster than \textsf{Stat} on directed and about 33000 times on undirected graphs (geometric means of the speedups). Since \textsf{SI} has shown to outperform other approaches in the context of updating the betweenness centrality of a single node after an edge insertion, we use it to update the betweenness in the greedy algorithm for the Maximum Betweenness Improvement problem (Section~\ref{sec:greedy}).
Therefore, in all the following experiments, what we refer to as \Greedy is the Algorithm of Section~\ref{sec:greedy} where we recompute betweenness after each edge insertion with \textsf{SI}.

\subsubsection{Running times of the greedy algorithm for betweenness maximization}
In Section~\ref{sec:greedy}, we already showed that \Greedy outperforms all other heuristics in terms of solution quality, both on directed and on undirected graphs (although we recall that the theoretical guarantee on the approximation ratio holds only for directed graphs). In this section, we report the running times of \Greedy, using \textsf{SI} to recompute betweenness. Table~\ref{table:times_heu_directed} and Table~\ref{table:times_heu_undirected} show the results on directed and undirected graphs, respectively. 
For each value of $k$, the tables show the running time required by \Greedy when $k$ edges are added to the graph. Notice that this is not the running time of the $k$th iteration, but the total running time of \Greedy for a certain value of $k$. 
Since on directed graphs the betweenness of $x$ is a submodular function of the solutions for \MBI (see Theorem~\ref{theo:submodular}), we can speed up the computation for $k>1$ (see Algorithm~\ref{alg:greedysub}). This technique was originally proposed in~\cite{M78} and it was used in~\cite{CDSV16} to speed up the greedy algorithm for harmonic centrality maximization.
Let $\Delta b_v(u) = b_v(S\cup\{(u,v)\}) -b_v(S)$, where $S$ is the solution computed at some iteration $i'<i$, that is, $\Delta b_v(u)$ is the increment to $b_v$ given by adding the edge $(u,v)$ to $S$ at iteration $i'$. Let $LB$ be the current best solution at iteration $i$. We avoid to compute $b_v(S\cup\{(u,v)\})$ at line \ref{greedy6} if $LB \geq b_v(S) + \Delta b_v(u)$. In fact, by definition of submodularity, $\Delta b_v(u)$ is monotonically non-increasing and $b_v(S) + \Delta b_v(u)$ is an upper bound for $b_v(S\cup\{(u,v)\})$. Then, $LB \geq b_v(S) + \Delta b_v(u)$ implies $LB \geq b_v(S\cup\{(u,v)\})$.

\label{sec:greedysub}
\begin{algorithm2e}[t]
\caption{\Greedy algorithm with pruning (exploiting submodularity).}
\label{alg:greedysub}
\SetKwInput{Proc}{Algorithm}
\SetKwInOut{Input}{Input}
\SetKwInOut{Output}{Output}
\Input{A directed graph $G=(V,E)$; a vertex $v\in V$; and an integer $k\in\mathbb{N}$}
\Output{Set of edges $S\subseteq \{(u,v)~|~u\in V\setminus N_v\}$ such that $|S|\leq k$}
$S\leftarrow\emptyset$\;
\ForEach{$u\in V\setminus( N_v(S))$}
{ 
$\Delta b_v(u) \leftarrow 0$\;
}
\For{$i=1,2,\ldots,k$} 
{\label{greedy4}
$LB \leftarrow 0$\;
\ForEach{$u\in V\setminus( N_v(S))$}
{ \label{greedy5}
 \If{$ (i=1)\vee (LB < (b_v(S) + \Delta b_v))$}
   { 
Compute $b_v(S\cup\{(u,v)\})$\; \label{greedy6}
$\Delta b_v(u) \leftarrow b_v(S\cup\{(u,v)\}) -b_v(S)$\;
$LB \leftarrow \text{max}(LB, b_v(S\cup\{(u,v)\}))$\;
   }\label{alg:dyngreedy:if}

} 
$u_{\max} \leftarrow \arg\max\{b_v(S\cup\{(u,v)\}) ~|~u\in V\setminus( N_v(S))\}$\;
$S \leftarrow S \cup \{(u_{\max},v)\}$\;

}
\Return $S$\;
\end{algorithm2e}

Although the standard deviation is quite high, we can clearly see that exploiting submodularity has significant effects on the running times: for all graphs in Table~\ref{table:times_heu_directed}
, we see that the difference in running time between computing the solution for $k= 1$ and $k = 10$ is at most a few seconds. Also, for all graphs the computation never takes more than a few minutes.
\begin{table}[h]
\caption{The left part of the table reports the running times (in seconds) of \Greedy on directed real-world graphs for different values of $k$. The right part shows the standard deviations.}
\begin{scriptsize}
  \begin{tabular}{ | l | r | r | r | r | r | r | r | r |}
  \cline{2-9}
      \multicolumn{1}{ l |}{} & \multicolumn{4}{| c |}{Running time \textsf{Greedy}} & \multicolumn{4}{| c |}{STD. DEV. \textsf{Greedy}}\\
      \hline
Graph  &  $k = 1$  &  $k = 2$  &  $k = 5$  & $k = 10$  &  $k = 1$  &  $k = 2$  &  $k = 5$  &  $k = 10$  \\
\hline
\texttt{subelj-jung}  &  1.79  & 1.91  & 1.99  & 2.10  &  0.56  & 0.58  & 0.61  & 0.68  \\
\texttt{wiki-Vote}  &  14.32  & 14.44  & 14.74  & 15.19  &  10.75  & 10.81  & 11.04  & 11.46  \\
\texttt{elec}  &  12.47  & 12.57  & 12.81  & 13.13  &  7.80  & 7.83  & 7.99  & 8.16  \\
\texttt{freeassoc}  &  81.52  & 83.01  & 87.00  & 96.60  &  66.27  & 67.88  & 70.84  & 82.01  \\
\texttt{dblp-cite}  &  584.90  & 694.19  & 710.90  & 729.73  &  1060.50  & 1268.18  & 1296.99  & 1328.83  \\
\texttt{subelj-cora}  &  1473.04  & 1504.96  & 1600.68  & 1688.39  &  1491.48  & 1526.95  & 1657.98  & 1784.74  \\
\texttt{ego-twitter}  &  164.43  & 179.13  & 217.19  & 229.39  &  200.10  & 211.52  & 259.85  & 275.22  \\
\texttt{ego-gplus}  &  211.39  & 225.58  & 230.26  & 240.29  &  195.22  & 186.00  & 188.82  & 196.78  \\
\texttt{munmun-digg}  &  736.13  & 739.82  & 749.74  & 759.58  &  313.45  & 313.50  & 313.66  & 316.35  \\
\texttt{linux}  &  1145.94  & 1239.16  & 1271.74  & 1311.28  &  822.06  & 917.50  & 933.02  & 951.61  \\

\hline
  \end{tabular}
  \end{scriptsize}
 \label{table:times_heu_directed}
\end{table}
\begin{table}[h]
\caption{The left part of the table reports the running times (in seconds) of \Greedy on undirected real-world graphs for different values of $k$. The right part shows the standard deviations.}
\begin{scriptsize}
  \begin{tabular}{ | l | r | r | r | r | r | r | r | r | r | r |}
  \cline{2-9}
      \multicolumn{1}{ l |}{} & \multicolumn{4}{| c |}{Running time \textsf{Greedy}} & \multicolumn{4}{| c |}{STD. DEV. \textsf{Greedy}}\\
      \hline
      Graph  &  $k = 1$  &  $k = 2$  &  $k = 5$  & $k = 10$  &  $k = 1$  &  $k = 2$  &  $k = 5$  &  $k = 10$  \\
\hline
\texttt{Mus-musculus}  &  27.06  & 87.80  & 394.30  & 1155.46  &  15.53  & 36.35  & 176.05  & 630.80  \\
\texttt{HC-BIOGRID.}  &  34.54  & 85.98  & 289.84  & 701.50  &  9.63  & 25.63  & 100.04  & 217.76  \\
\texttt{Caenor-elegans}  &  11.17  & 25.47  & 94.94  & 320.85  &  3.23  & 10.23  & 23.66  & 55.19  \\
\texttt{ca-GrQc.}  &  19.76  & 43.01  & 149.43  & 438.98  &  8.64  & 20.65  & 53.63  & 96.31  \\
\texttt{advogato}  &  12.42  & 28.07  & 81.79  & 299.05  &  1.56  & 13.96  & 28.23  & 147.66  \\
\texttt{hprd-pp}  &  47.08  & 111.85  & 460.31  & 1561.82  &  12.84  & 29.65  & 59.01  & 439.32  \\
\texttt{ca-HepTh}  &  100.34  & 464.66  & 2069.34  & 5926.75  &  42.83  & 282.09  & 604.61  & 1320.20  \\
\texttt{dr-melanog}  &  71.43  & 160.89  & 614.92  & 2084.71  &  18.01  & 31.55  & 46.88  & 333.84  \\
\texttt{oregon1}  &  30.66  & 69.06  & 191.63  & 441.09  &  4.87  & 9.41  & 23.99  & 76.51  \\
\texttt{oregon2}  &  36.44  & 73.35  & 233.28  & 594.53  &  9.63  & 16.92  & 25.26  & 44.3 7  \\
\texttt{Homo-sapiens}  &  99.82  & 276.09  & 1155.97  & 3554.53  &  20.30  & 54.42  & 258.89  & 673.55  \\
\texttt{GoogleNw}  &  68.33  & 102.35  & 220.32  & 451.29  &  11.71  & 17.18  & 36.19  & 76.37  \\
\texttt{CA-CondMat}  &  1506.68  & 3402.10  & 12177.24  & 36000.24  &  381.00  & 927.40  & 2178.47  & 17964.74  \\

\hline
  \end{tabular}
  \end{scriptsize}

  \label{table:times_heu_undirected}
\end{table}

Unfortunately, submodularity does not hold for undirected graphs, therefore for each $k$ we need to apply \textsf{SI} to all possible new edges between $x$ and other nodes. Nevertheless, apart from the \texttt{CA-CondMat} graph (where, on average, it takes about 10 hours for $k = 10$) and \texttt{ca-HepTh} (where it takes about 1.5 hours), \Greedy never requires more than 1 hour for $k = 10$. For $k = 1$, it takes at most a few minutes. Quite surprisingly, the running time of the first iteration is often smaller than that of the following ones, in particular if we consider that the first iteration also includes the initialization of \textsf{SI}. This might be due to the fact that, initially, the pivots are not very central and therefore many edge insertions between the pivots and other nodes affect only a few shortest paths. Since the running time of \textsf{IA} is proportional to the number of affected node pairs, this makes it very fast during the first iteration. On the other hand, at each iteration the pivot $x$ gets more and more central, affecting a greater number of nodes when a new shortcut going through $x$ is created.
 
To summarize, our experimental results show that our incremental algorithm for the betweenness of one node is much faster than existing incremental algorithms for the betweenness of all nodes, taking always fractions of seconds even when the competitor takes seconds. The combination of it with our greedy approach for the \MBI problem allows us to maximize betweenness of graphs with hundreds of thousands of edges in reasonable time. Also, our results in Section~\ref{sec:experiments_greedy} show that \Greedy outperforms other heuristics both on directed and undirected graphs and both for the problem of betweenness and ranking maximization.

\section{Conclusions}
\label{sec:conclusions}
Betweenness centrality is a widely-used metric that ranks the importance of nodes in a network. 
Since in several scenarios a high centrality directly translates to some profit, in this paper we have studied the problem of maximizing the betweenness of a vertex by inserting a predetermined number of new edges incident to it. Our greedy algorithm, which is a $(1 - \frac{1}{e})-$approximation of the optimum for directed graphs, yields betweenness scores that are significantly higher than several other heuristics, both on directed and undirected graphs. Our results are drawn from experiments on a diverse set of real-world directed and undirected networks with up to $10^5$ edges. 

Also, combining our greedy approach with a new incremental algorithm for recomputing the betweenness of a node after an edge insertion, we are often able to find a solution in a matter of seconds or few minutes. Our new incremental algorithm extends a recently published APSP algorithm and is the first to recompute the betweenness of one node in $O(n^2)$ time. All other existing approaches recompute the betweenness of all nodes and require at least $O(nm)$ time, matching the worst-case complexity of the static algorithm. 
Although extremely fast, our betweenness update algorithm has a memory footprint of $\Theta(n^2)$, which is a limitation for very large networks. 
A possible direction for future work could be to combine our greedy approach with dynamic algorithms that compute an approximation of betweenness centrality. Since these algorithms require less memory than the exact ones, they might allow us to target even larger networks. 

Also, future work could consider extensions of the problem studied in this paper, such as allowing additions of edges incident to other vertices or weight changes to the existing edges.



\begin{acks}
This work is partially supported by German Research Foundation (DFG) grant ME-3619/3-2 (FINCA) within the Priority Programme 1736 \emph{Algorithms for Big Data} and by the Italian Ministry of Education, University, and Research (MIUR) under PRIN 2012C4E3KT national research project AMANDA - Algorithmics for MAssive and Networked DAta.
We thank Dominik Kiefer (Karlsruhe Institute of Technology) for help with the experimental evaluation. 
\end{acks}

\bibliographystyle{ACM-Reference-Format-Journals}
\bibliography{references}

\end{document}